%% file: main.tex
\documentclass[fleqn,usenatbib]{mnras}

\usepackage{newtxtext,newtxmath}

\usepackage[T1]{fontenc}

\DeclareRobustCommand{\VAN}[3]{#2}
\let\VANthebibliography\thebibliography
\def\thebibliography{\DeclareRobustCommand{\VAN}[3]{##3}\VANthebibliography}

\usepackage{xcolor}
\usepackage[normalem]{ulem}

\usepackage{xspace}
\newcommand{\NpreResolved}{101\xspace}
\newcommand{\NpreBlended}{40\xspace}
\newcommand{\NpreOneArcsec}{141\xspace}
\newcommand{\NpreImaging}{337\xspace}
\newcommand{\NpreImagingTwo}{102\xspace}
\newcommand{\NpreImagingOne}{235\xspace}
\newcommand{\NpreShared}{36\xspace}
\newcommand{\NpreTotal}{442\xspace}
\newcommand{\NconfirmedDual}{9\xspace}
\newcommand{\NlitPairVerified}{6\xspace}
\newcommand{\NlitGeometryVerified}{2\xspace}
\newcommand{\Nlens}{26\xspace}
\newcommand{\Nstar}{4\xspace}
\newcommand{\Ngal}{0\xspace}
\newcommand{\Nproj}{1\xspace}
\newcommand{\Nnonlens}{5\xspace}
\newcommand{\Nartefact}{1\xspace}
\newcommand{\Nresolved}{98\xspace}
\newcommand{\Nblended}{38\xspace}
\newcommand{\Nimaging}{335\xspace}
\newcommand{\NimagingOne}{235\xspace}
\newcommand{\NimagingTwo}{100\xspace}
\newcommand{\Nshared}{35\xspace}
\newcommand{\Nfinal}{436\xspace}
\newcommand{\NmissingPM}{113\xspace}
\newcommand{\NresolvedAboveThreshold}{137\xspace}
\newcommand{\SepRangeAll}{$0.177$--$1.498$}
\newcommand{\SepRange}{\SepRangeAll$^{\prime\prime}$\xspace}
\newcommand{\zRangeAll}{$0.166$--$3.769$}
\newcommand{\zRange}{\zRangeAll\xspace}
\newcommand{\kpcRangeAll}{$1.50$--$12.99$}
\newcommand{\kpcRange}{\kpcRangeAll\,kpc\xspace}
\newcommand{\LnRthreshold}{$-6.81$\xspace}
\newcommand{\CalibN}{109\xspace}
\newcommand{\CalibCompleteness}{87.1\%\xspace}
\newcommand{\CalibPurity}{66.0\%\xspace}
\newcommand{\CalibInterval}{$-7.83\leq t\leq-2.58$\xspace}
\newcommand{\Ncontrolfields}{60,000\xspace}
\newcommand{\BchanceResolved}{21.3\xspace}
\newcommand{\BchanceBlended}{23.6\xspace}
\newcommand{\BchanceTwo}{43.1\xspace}
\newcommand{\BchanceOne}{283.4\xspace}

\newcommand{\zRangeRes}{0.387--3.082\xspace}
\newcommand{\zRangeBl}{0.166--3.673}

\newcommand{\zMedRes}{1.524}
\newcommand{\zMedBl}{1.080}

\newcommand{\SepRangeRes}{0.421--0.998}
\newcommand{\SepRangeBl}{0.311--0.997}

\newcommand{\kpcRangeRes}{2.76--8.52}
\newcommand{\kpcRangeBl}{1.95--8.64}

\newcommand{\kpcMedRes}{6.45}
\newcommand{\kpcMedBl}{5.24}

\newcommand{\NlitMatchRes}{6\xspace}
\newcommand{\NlitMatchBl}{0}

\newcommand{\NimagingPreDedupOne}{1139\xspace}
\newcommand{\NimagingPreDedupTwo}{491\xspace}
\newcommand{\NimagingScored}{1600\xspace}
\newcommand{\NimagingEligible}{607\xspace}
\newcommand{\NscoreImaging}{406\xspace}
\newcommand{\NscoreImagingOne}{238\xspace}
\newcommand{\NscoreImagingTwo}{168\xspace}
\newcommand{\NpreMorphImaging}{340\xspace}
\newcommand{\NpreMorphImagingOne}{238\xspace}
\newcommand{\NpreMorphImagingTwo}{102\xspace}
\newcommand{\NresolvedEligibleMagdep}{172\xspace}
\newcommand{\NresolvedEligibleMagindep}{190\xspace}
\newcommand{\NresolvedScoreMagindep}{135\xspace}
\newcommand{\NresolvedScoreMagdep}{123\xspace}
\newcommand{\NblendedFinite}{140\xspace}
\newcommand{\NblendedQQMagdep}{36\xspace}
\newcommand{\NblendedQQMagindep}{36\xspace}
\newcommand{\NblendedQQBoth}{27\xspace}
\newcommand{\NblendedQQEither}{45\xspace}
\newcommand{\NblendedLegacyR}{18\xspace}
\newcommand{\NblendedFallback}{20\xspace}
\newcommand{\NblendedPSone}{8\xspace}
\newcommand{\NblendedNSC}{12\xspace}
\newcommand{\NlitDuals}{55\xspace}
\newcommand{\NlitUnmatched}{46\xspace}
\newcommand{\NlitNotInMaster}{18\xspace}
\newcommand{\NlitIndependent}{8\xspace}
\newcommand{\NlitIndependentImaged}{4\xspace}
\newcommand{\NlitNoSecondary}{17\xspace}
\newcommand{\NlitFailPSF}{6\xspace}
\newcommand{\NlitFailCuts}{5\xspace}
\newcommand{\NeuclidCovered}{44\xspace}
\newcommand{\NeuclidVIS}{40\xspace}
\newcommand{\NeuclidTwo}{39\xspace}
\newcommand{\NeuclidBlend}{1\xspace}
\newcommand{\NeuclidResolved}{8\xspace}
\newcommand{\NeuclidBlended}{3\xspace}
\newcommand{\NeuclidImaging}{31\xspace}
\newcommand{\NeuclidShared}{2\xspace}
\newcommand{\NeuclidMain}{17\xspace}
\newcommand{\NeuclidAdditional}{23\xspace}

\newcommand{\zRangeImTwo}{0.594--3.187}
\newcommand{\zMedImTwo}{1.677}
\newcommand{\SepRangeImTwo}{0.475--1.490}
\newcommand{\kpcRangeImTwo}{4.12--12.92}
\newcommand{\kpcMedImTwo}{9.54}
\newcommand{\NlitMatchImTwo}{4}
\newcommand{\zRangeMain}{0.166--3.673}
\newcommand{\zMedMain}{1.525}
\newcommand{\SepRangeMain}{0.311--1.490}
\newcommand{\kpcRangeMain}{1.95--12.92}
\newcommand{\kpcMedMain}{6.98}
\newcommand{\NlitMatchMain}{9}
\newcommand{\zRangeOneGaia}{0.435--3.769}
\newcommand{\zMedOneGaia}{1.639}
\newcommand{\SepRangeOneGaia}{0.177--1.498}
\newcommand{\kpcRangeOneGaia}{1.50--12.99}
\newcommand{\kpcMedOneGaia}{10.74}
\newcommand{\NlitMatchOneGaia}{0}
\newcommand{\Nmain}{201\xspace}
\newcommand{\Bmain}{88\xspace}

\usepackage{graphicx}	%
\usepackage{amsmath}	%
\usepackage{float}
\usepackage{placeins}
\usepackage{subcaption}
\usepackage{tikz}
\usetikzlibrary{arrows.meta,positioning}
\usepackage{booktabs}
\usepackage{tabularx}
\usepackage{float}
\usepackage{booktabs}
\usepackage{longtable}
\usepackage{pdflscape}
\usepackage{lastpage}

\title[Hunting arcsecond-scale dual and lensed quasars]{Hunting arcsecond-scale dual and lensed quasars in wide-field surveys}

\author[L. M. Cooper et al.]{
Lewis M. Cooper,$^{1}$\thanks{E-mail: lmc210@cam.ac.uk}
Vasily Belokurov,$^{1}$
and Sergey E. Koposov$^{2,1}$
\\
$^{1}$Institute of Astronomy, University of Cambridge, Madingley Road, Cambridge, CB3 0HA, UK\\
$^{2}$Institute for Astronomy, University of Edinburgh, Royal Observatory, Blackford Hill, Edinburgh EH9 3HJ, UK
}

\date{Accepted XXX. Received YYY; in original form ZZZ}

\pubyear{\the\year{}}

\begin{document}
\label{firstpage}
\pagerange{\pageref{firstpage}--\pageref{LastPage}}
\maketitle

\begin{abstract}
Dual active galactic nuclei (AGN) at kiloparsec separations probe the coevolution between galaxies and their central supermassive black holes (SMBHs). They are potential progenitors of compact SMBH binaries, but survey resolution limits and chance projections complicate their identification. We present two complementary pipelines to identify dual and lensed quasar candidates in Gaia-matched quasars from the SDSS, DESI and Quaia catalogues using ground-based imaging and Gaia source multiplicity. Both searches assess photometric consistency with two quasars at the adopted system redshift. The Gaia search yields \Nresolved resolved and \Nblended blended candidates, and the imaging search yields \Nimaging, of which \NimagingOne have one Gaia detection while both searches share \Nshared systems. We divide the candidates into a main tier of \Nmain systems with two distinct Gaia detections and an additional tier of \NimagingOne systems with one Gaia detection. Of these, \NconfirmedDual are confirmed dual AGN and \Nlens are known lensed quasars, all in the main tier. To measure how often unrelated sources pass our selection, we score sources from \Ncontrolfields random quasar fields as fake companions. At most \Bmain of the main-tier candidates are expected to be chance alignments, while the one-Gaia candidates are consistent with chance alignments and are provided as targets for follow-up. The retained candidates span angular separations of \SepRange and redshifts of \zRange. Assuming that both components share the adopted redshift, their projected separations span \kpcRange. The catalogue provides targets for spatially resolved spectroscopy and high-resolution imaging to distinguish dual AGN from lenses.
\end{abstract}


\section{Introduction}
The hierarchical paradigm of galaxy formation predicts that mergers are frequent events in galactic evolution \citep{begelmanMassiveBlackHole1980, kormendyInwardBoundTheSearch1995, richstoneSupermassiveBlackHoles1998, milosavljevicFormationGalacticNuclei2001, yuEvolutionMassiveBinary2002} and significantly influences galactic morphologies and star formation rates \citep{whiteCoreCondensationHeavy1978}. Most massive galaxies are also thought to ubiquitously host a supermassive black hole (SMBH) with a typical mass between $10^6\mathrm{M_\odot}$ to $10^{10}\mathrm{M_\odot}$ at their centres \citep{shankarProbingBlackHole2020, greeneIntermediateMassBlackHoles2020}. A key outcome of galaxy mergers is the formation of dual SMBHs resulting from the dynamical friction between the two dark matter halos leading to the SMBHs drifting towards one another \citep{mayerRapidFormationSupermassive2007, kormendyCoevolutionNotSupermassive2013a}. Subsequent interactions with surrounding material and gravitational torques may further reduce the binary separation, eventually allowing gravitational-wave emission to dominate its orbital evolution and drive the final inspiral and coalescence \citep{volonteriAssemblyMergingHistory2003, callegariPairingSupermassiveBlack2009, krauseEvidenceSupermassiveBlack2025}.

Hierarchical models and hydrodynamical simulations suggest that the galaxy mergers drives gas towards the central SMBHs, potentially triggering active galactic nucleus (AGN) activity. The system is labelled a dual AGN when both SMBHs are actively accreting matter simultaneously \citep{laceyMergerRatesHierarchical1994,capeloGrowthActivityBlack2015,saeedzadehDualActiveGalactic2024}. Dual AGN at kpc scale separations are therefore expected as a natural stage in the evolutionary sequence from merging galaxies to a bound SMBH binary \citep{mayerRapidFormationSupermassive2007,saeedzadehDualActiveGalactic2024}.
Such systems provide some of the only direct electromagnetic evidence for the coevolution of SMBHs with their host galaxies \citep{chenCloseQuasarPair2023,suhRapidGrowthDual2026}.
However, at cosmological distances, such systems remain difficult to detect observationally because their arcsecond-scale angular separations are difficult to resolve with current surveys \citep{mannucciUnveilingPopulationDual2022a,shenStatisticsGalacticscaleQuasar2023}.

Although a dual AGN consists of two physically associated active nuclei, there is no standard observational definition. Some studies restrict the separation to less than $10\,\mathrm{kpc}$, while others extend the separation limit to $50\mathrm{kpc}$ and require that the line-of-sight velocity separation satisfy $\Delta v< 600 \mathrm{km\,s^{-1}}$ \citep{comerfordDUALSUPERMASSIVEBLACK2013, dadianiCosmicPairsDESI2026}. In cases where both objects are members of the most luminous population of AGN, quasars (QSOs), at $z > 0.5$, studies often extend the definition to $\Delta v< 2000 \mathrm{km\,s^{-1}}$ to account for systematic uncertainties of order $1000 \mathrm{km\,s^{-1}}$ in broad UV emission line redshifts \citep{hennawiBinaryQuasarsSloan2006, hennawiBINARYQUASARSHIGH2010}.

At sub-parsec separations, a binary SMBH may be detectable through periodic light curve variability or low frequency gravitational waves \citep{charisiMultimessengerAnalysisSupermassive2025}. However, the systematic identification of such systems prior to the parsec separation stage (at kpc separations) remains observationally challenging with current electromagnetic observatories and a trade-off often exists between wide-field, all sky searches and precise, high resolution observations of individual systems. The search for dual AGN can be aided by targeting luminous quasars during the epoch of their greatest abundance, Cosmic noon (approximately $1\lesssim z\lesssim3$), when cosmic star formation, SMBH accretion and merger activity peaked \citep{schmidtSpaceDistributionLuminosity1968, laceyMergerRatesHierarchical1994, veronCosmicEvolutionQuasars1986, shaverRedshiftCutoffQuasar1999}. Together with frequent galaxy mergers, these conditions make this cosmic period particularly relevant for investigating the abundance of dual AGN. Quasars therefore provide luminous observational signposts of active SMBH growth and, through their connection to galaxy evolution, help trace the coevolution between SMBH and host galaxy across cosmic time \citep{kormendyCoevolutionNotSupermassive2013a, beckmannMassiveBlackHoles2026a}. Establishing the abundance and separation distribution of dual AGN across cosmic time can consequently constrain the processes governing SMBH pairing and the triggering of simultaneous accretion during galaxy mergers. Comparisons with theoretical predictions can help connect the observable kpc scale population to the evolutionary pathways leading to bound SMBH binaries. 

Despite their significance, identifying dual AGN on kpc scales remains observationally challenging, and such systems confirmed through high resolution imaging or resolved spectroscopy are outnumbered by the number of candidate systems by orders of magnitude \citep[e.g.][]{pfeifleSupersizeMeBig2025}. The search for dual AGNs has historically been limited by a number of factors including angular resolution and deblending limits of optical surveys, the prevalence of common contaminants, and the inherent scarcity of such systems. Quasars, named for their "quasi-stellar" appearance \citep{seyfertNuclearEmissionSpiral1943}, resemble point source, stellar objects and can be observationally conflated with stars. The identification of individual quasars can overcome this issue with spectroscopic or photometric observations to detect the extreme luminosity, non-thermal continuum emission, strong broad emission lines, and/or variability on short timescales typical of AGN \citep{mushotzkyXraySpectraTime1993, vandenberkCompositeQuasarSpectra2001, richardsSpectralEnergyDistributions2006, shenCatalogQuasarProperties2011, wuCatalogQuasarProperties2022}. Identifying dual AGNs, however, is more challenging since many signatures of two compact, largely featureless point-like objects at small separations are not unique to AGN. Foreground stars and unresolved substructure may appear as a projected companion to the quasar which can be difficult to isolate spectroscopically, particularly when the flux ratio of the system is dominated by a quasar or the projected separation is small enough for the system to appear partially \citep{mukhejaDecouplingCandidateDual2026}. Additionally, the prevalence of gravitationally lensed quasars can also mimic dual AGN, often requiring deep, high-resolution imaging to distinguish lensed quasars from true dual AGN systems. Late-stage mergers are also often heavily obscured \citep{ricciGrowingSupermassiveBlack2017, kossPopulationLuminousAccreting2018}, which can hide one nucleus even at sufficient resolution. Wide separation systems, therefore, dominate the currently known population and leave much of the arcsecond scale regime unknown.

Estimates for the fraction of dual AGN systems vary depending on the separation and redshift regimes, observational techniques and selection methods. Recent estimates of SDSS quasars range from $\sim0.06\%$ to $0.26\pm0.18\%$ at $\lesssim 30\mathrm{kpc}$ scales \citep{silvermanDualSupermassiveBlack2020, tangSpectroscopicConfirmationDual2026}. On the sub-arcsecond regime, {\citet{shenStatisticsGalacticscaleQuasar2023} measured a double quasar fraction of $0.062\pm0.005\%$ across $0.3-3^{\prime\prime}$ at $1.5<z<3.5$, however this combines physical pairs and gravitational lenses, and therefore should be interpreted as a double quasar fraction.

Despite this, several approaches searching for dual AGN on a range of separation scales, including arcsecond scales, have expanded our knowledge of SMBH pairing. Early searches exploited the double-peaked emission lines in high resolution spectra which are associated with two dynamic, independent nuclei \citep{shiSearchDoublepeakedNarrow2014a, maschmannDoublepeakEmissionLine2020, zhengSDSSJ15323180+4203427Triple2026}. When the narrow line regions of two merging AGN are well separated in velocity, two sets of strong emission lines (e.g. $\mathrm{[OIII]}, \mathrm{H}\beta $) can often be detected with sufficient spectroscopic resolution \citep{blechaDoublepeakedNarrowlineSignatures2013, krauseEvidenceSupermassiveBlack2025}. However, because double-peaked emission lines can also arise from other processes, including outflows and rotating discs \citep[e.g.,][]{muller-sanchezORIGINDOUBLEPEAKEDNARROW2015, comerfordOriginDoublepeakedNarrow2018, maschmannDoublepeakEmissionLine2020, prajnadiptOriginDoublepeakedNarrow2026}, their use as a stand-alone detection for binary AGN is prone to contamination. 

Several recent searches have shifted the study of close binary quasars and dual AGN from individual discoveries toward systematic, survey-based selection of compact dual and binary AGN candidates. 

Large, wide-field surveys such as SDSS, DESI and Gaia now allow such searches over millions of quasars. One such example of these approaches is the exploitation of time-dependent astrometric perturbations from unresolved dual AGN in Gaia observations through the Varstrometry for Off-nucleus and Dual sub-Kpc AGN (VODKA) programme \citep{shenVarstrometryOffnucleusDual2019}. Intrinsic AGN variability changes the flux ratio between unresolved components, causing the measured photocentre to shift between epochs such that the resulting astrometric jitter can appear as excess astrometric noise or poor single-source astrometric fits. In its initial implementation using Gaia DR2, VODKA was applied to $\sim86,000$ SDSS quasars, identifying a large population of sources with elevated astrometric excess noise and suggesting that at least hundreds of sub-kpc dual AGN candidates should be accessible to follow-up \citep{hwangVarstrometryOffnucleusDual2020}. Follow-up observations from the Hubble Space Telescope (HST) then tested this selection directly on a sample of 84 systems, of which 45 were resolved into multiple components, including 19 systems interpreted as likely physical quasar pairs or gravitationally lensed quasars, alongside substantial contamination from quasar-star chance alignments \citep{chenVarstrometryOffnucleusDual2022}. In Gaia DR3, VODKA was applied to 134,796 SDSS DR16Q quasars with Gaia sources within $3\arcsec$ at $z>1.5$. The final cleaned catalogue contains 60 double quasar candidates with separations of $\sim0.3-3^{\prime\prime}$ \citep{shenStatisticsGalacticscaleQuasar2023}. Follow-up observations of a small separation subset has so far classified nine systems, finding two dual AGN and three quadruply lensed quasars \citep{baltaciVarstrometryOffnucleusDual2026}. Astrometric diagnostics efficiently identify multiplicity, but high-resolution imaging and spectroscopy remain necessary to distinguish dual AGN from lenses.

A complementary method was provided by the Gaia multipeak method (GMP), which uses the DR3 image parameter determination (IPD) statistic, \texttt{ipd\_frac\_multi\_peak}, to select sources which are inconsistent with a single compact object. This approach is particularly relevant to the sub-arcsecond regime because it can select candidates below the normal resolution limit of wide-area imaging surveys. \citet{ciurloNewMultipleAGN2023} observed four GMP candidates and found that all four were resolved, with two confirmed dual AGN, one dual/triple-AGN or lensed-dual system, and one AGN-star alignment. \citet{mannucciGMPselectedDualLensed2023} further quantified the selection function, showing that GMP is effective for separations $\gtrsim 0.15\arcsec$ when the secondary source is sufficiently bright, and reported a sample of 22 systems comprising 10 dual AGN, 5 lensed systems, and 7 AGN-star chance alignments. These targeted searches sit within the broader census provided by the Big Multi-AGN Catalog (BigMAC), which compiles 156 confirmed dual AGN, 4,180 dual-AGN candidates, one confirmed gravitationally bound binary AGN and 1,368 binary-AGN/SMBH candidates from the literature \citep{pfeifleSupersizeMeBig2025}. Most recently, \citet{zanchettinDualLensedAGN2026} performed high resolution spectroscopic follow-up of 34 sub-arcsecond multi-AGN candidates selected using the GMP method. Using the Enhanced Resolution Imager and Spectrograph (ERIS) at the ESO Very Large Telescope, they confirmed six dual AGN and seven gravitationally lensed quasars. Gaia multiplicity around quasars has also been used to search for gravitationally lensed quasars \citep{krone-martinsGaiaDR2Gravitational2018, lemonGravitationallyLensedQuasars2019, lemonGravitationallyLensedQuasars2023}, and these searches recover many of the same systems as dual-AGN searches.

In this study, we expand upon these searches and present two complementary pipelines for the identification of candidate dual AGN at arcsecond and sub-arcsecond separations, both drawing on a sample of photometrically and spectroscopically identified quasars. The first search identifies quasars with a neighbouring Gaia DR3 source within one arcsecond. The second uses DECaLS imaging to identify multiplicity in point sources near known quasars. In both searches, we use redshift-dependent probabilistic models of quasar colours to assess whether the photometry supports the presence of two quasars at the adopted redshift of the primary. Where separate component photometry is available, we evaluate the colours of the individual sources and where the ground-based photometry blends the two components, we compare the evidence for two quasars at a common redshift with that for a quasar and an unrelated point source. We use proper motion measurements, image diagnostics and quality flags where applicable to reduce contamination from foreground stars and imaging artefacts. The resulting catalogue provides targets for high-resolution imaging and spatially resolved spectroscopy to establish the nature of the individual components.

\section{Data}
Our search for dual candidates are performed in three complimentary quasar catalogues from the Sloan Digital Sky Survey (SDSS) data release 16 (DR16) \citep{lykeSloanDigitalSky2020}, the Dark Energy Spectroscopic Instrument \citep[DESI DR1, ][]{desicollaborationDataRelease12026} and version 1.0.0 of the Gaia-unWISE Quasar Catalog $G<20$ \citep[Quaia, ][]{storey-fisherQuaiaGaiaunWISEQuasar2024}. Together, these catalogues provide complementary quasar samples with different sky coverage and selection functions. We construct the parent inputs separately for the Gaia one arcsecond neighbour and imaging searches, as described below.

\subsection{Quasar samples} We constructed the parent quasar sample by combining SDSS DR16Q, DESI DR1, and Quaia $G<20$ catalogues. SDSS DR16Q contains 750,414 spectroscopically identified quasars observed across SDSS-I--IV, including 225,082 quasars appearing in an SDSS data release for the first time. We selected the 749,749 objects flagged as final, confident quasars ($\texttt{is\_qso\_final = 1}$). 

When compared with SDSS, DESI provides optical spectra and redshifts over a substantially larger and partly complementary target sample. From the DR1 catalogue, we selected 1,645,842 primary spectra classified as quasars (\texttt{spectype = QSO}) and having no reported redshift fitting warning (\texttt{zwarn = 0}).

Finally, we supplemented the spectroscopic surveys with the all-sky Quaia catalogue constructed from Gaia DR3 quasar candidates and mid-infrared photometry from unWISE. Quaia applies astrometric and colour cuts to reduce contamination and provides redshift estimates obtained using a model trained on SDSS spectroscopy. The full Quaia catalogue contains 1,295,502 sources at $G<20.5$, with a broad focus on coverage and completeness. However, we adopted the cleaner $G<20$ catalogue, which contains 755,850 sources with a greater focus on lower contamination and more reliable redshift estimates. This sample greatly extends the sky coverage of the spectroscopic catalogues outside the SDSS and DESI footprints.

\begin{figure}
      \centering
      \resizebox{\linewidth}{!}{%
          \begin{tikzpicture}[
              font=\small,
              node distance=8mm and 12mm,
              catalogue/.style={
                  draw,
                  rounded corners=2pt,
                  minimum width=38mm,
                  minimum height=12mm,
                  fill=blue!10,
                  font=\bfseries,
                  align=center
              },
              operation/.style={
                  draw,
                  rounded corners=2pt,
                  text width=42mm,
                  minimum height=17mm,
                  fill=gray!8,
                  align=center
              },
              final/.style={
                  draw,
                  rounded corners=2pt,
                  minimum width=65mm,
                  minimum height=13mm,
                  fill=orange!18,
                  font=\bfseries,
                  align=center
              },
              connection/.style={
                  thick,
                  draw=gray!75
              },
              arrow/.style={
                  connection,
                  -{Latex[length=2mm]}
              }
          ]

          \node[catalogue] (sdss)
              {SDSS DR16Q\\
               {\normalfont \textit{Selected inputs: 749,749}}};

          \node[catalogue, right=of sdss] (quaia)
              {Quaia $G<20$\\
               {\normalfont \textit{Selected inputs: 755,850}}};

          \node[catalogue, right=of quaia] (desi)
              {DESI DR1\\
               {\normalfont \textit{Selected inputs: 1,645,842}}};

          \node[operation, below=of sdss] (sdss-gaia)
              {Nearest Gaia DR3 source within
               $1^{\prime \prime}$\\
               \textit{489,328 matched}};

          \node[operation, below=of desi] (desi-gaia)
              {Nearest Gaia DR3 source within
               $1^{\prime \prime}$\\
               \textit{626,137 matched}};

          \coordinate (matching-level)
              at (quaia.south |- sdss-gaia.south);

          \node[
              circle,
              fill=gray!75,
              inner sep=1.7pt,
              below=12mm of matching-level
          ] (merge) {};

          \node[
              operation,
              text width=58mm,
              minimum height=21mm,
              below=8mm of merge
          ] (group)
              {Deduplication\\
               \textit{519,068 duplicate entries}};

          \node[final, below=of group] (master)
              {Master quasar catalogue\\
               \textit{1,352,247 unique entries}};

          \draw[arrow] (sdss) -- (sdss-gaia);
          \draw[arrow] (desi) -- (desi-gaia);

          \draw[connection] (sdss-gaia.south) -- (merge);
          \draw[connection] (desi-gaia.south) -- (merge);

          \draw[connection] (quaia.south) -- (merge);

          \draw[arrow] (merge) -- (group);
          \draw[arrow] (group) -- (master);

          \end{tikzpicture}%
      }
      \caption{Construction of the Gaia-linked master quasar catalogue from SDSS DR16Q, DESI DR1 and Quaia G<20. SDSS and DESI were matched to the nearest Gaia DR3 source within $1^{\prime \prime}$. 1,352,247 unique Gaia-linked quasars are present in the final master catalogue.}
      \label{fig:master-quasar-parent}
  \end{figure}
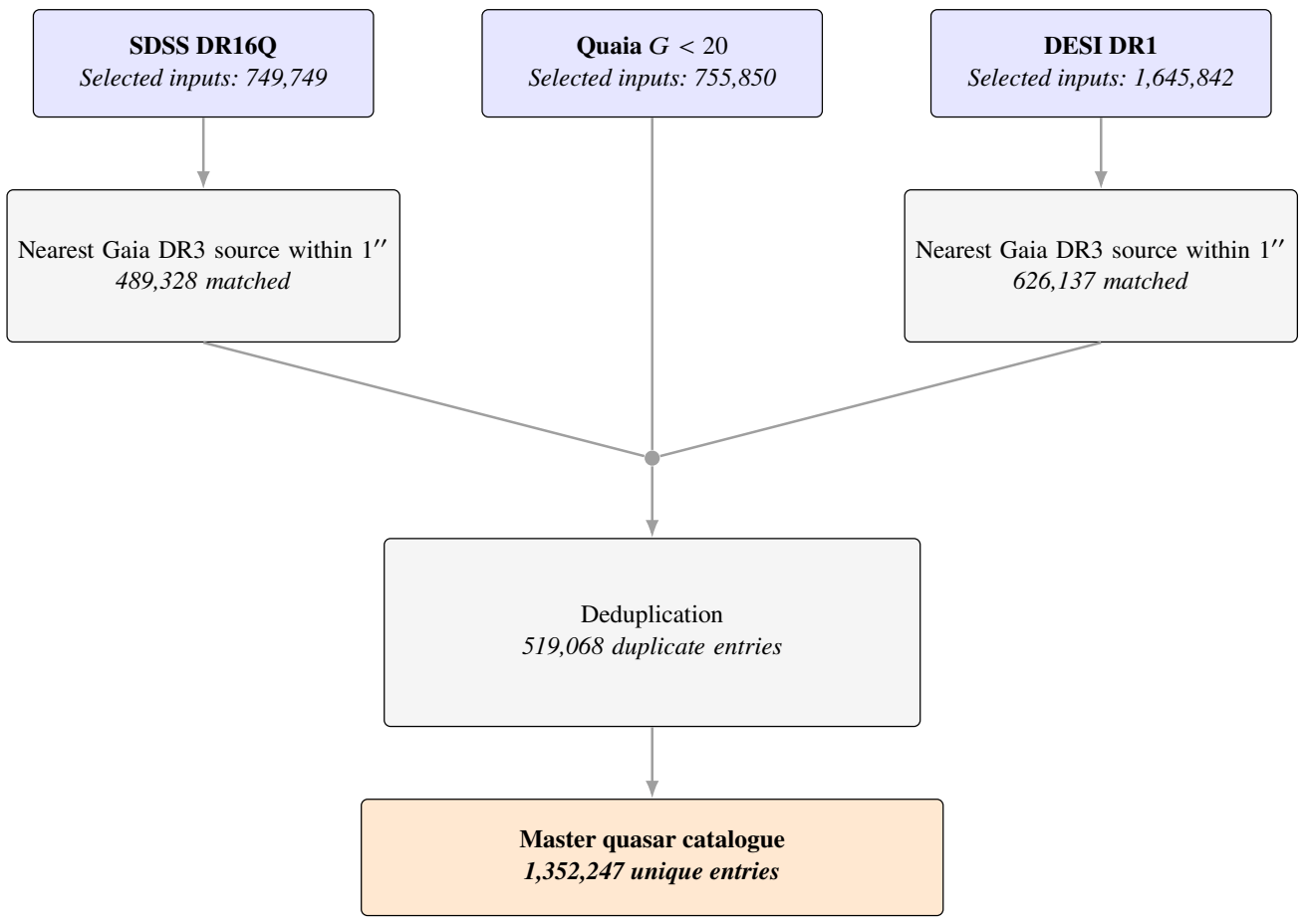

For the imaging search, we combine 749,749 SDSS DR16Q objects with \texttt{is\_qso\_final=1}, 1,645,842 primary DESI DR1 quasar spectra with \texttt{zwarn=0}, and 755,850 Quaia sources with $G<20$, giving 3,151,441 input records. Of these, 489,328 SDSS records and 626,137 DESI records are matched to their nearest Gaia DR3 source within $1^{\prime\prime}$. This gives 1,871,315 Gaia-linked records. Merging records that share a Gaia source identifier removes 519,068 duplicates, yielding 1,352,247 unique Gaia-linked parent entries. Where catalogues overlap, the imaging master adopts redshifts in the order of preference DESI, SDSS, then Quaia.

\subsection{Photometry, astrometry and imaging}
We make use of photometric colours and ancillary astrometric diagnostics to further refine our search. In particular we use Gaia DR3 crossmatched with our quasar catalogue specifically using the one arcsecond neighbour count quantity and proper motion. During subsequent colour analysis, we use supporting data from SDSS, ALLWISE \citep{wrightWidefieldInfraredSurvey2010}, PS1 \citep{chambersAstrometryPanSTARRSPS12005, flewellingPanSTARRS1DatabaseData2020}, NSC \citep{nideverSecondDataRelease2021}, SkyMapper \citep{onkenSkyMapperSouthernSurvey2024} and VHS \citep{mcmahonFirstScientificResults2013} for broader photometric coverage.

Our image analysis method relies on wide field Dark Energy Camera Legacy Survey \citep[DECaLS, ][]{deyOverviewDESILegacy2019} DR9 imaging data with $g,r,z$ photometry. We use the DECaLS Tractor model's point spread function (PSF) profiles and image, model and residual cutouts for our analysis.

\section{Candidate selection methods}
We present two supporting methods in the search for close separation dual quasars. The first investigates higher multiplicity components within 1 arcsecond of known quasars and the second explores non-singular PSF morphology in observed wide-field DECaLS imaging of known primary quasars and quasar-like additional components.

Both methods further assess the colour properties of each system or, where component photometry is available, of its individual components, based on the hypothesis from \citet{templeModellingType12021} that quasars occupy a narrow region of colour-redshift space owing to their self-similar spectra at fixed redshift. Variation in quasar colours across redshift can be explained by prominent emission lines entering or exiting specific colour bands.

\subsection{One arcsecond search}
Our first method searches for and classifies objects within a $1^{\prime \prime}$ radius of known Gaia-observed quasars. The Gaia-neighbour search uses a separately assembled sample selected from quasars associated with Gaia sources having \texttt{cnt1>1} in the DR3 auxiliary neighbour count table. This yields 712 input entries spanning $0.066\lesssim z\lesssim4.436$, with 114 assigned to SDSS, 240 to DESI and 358 to Quaia. Redshift eligibility is assessed subsequently by the scoring models. Gaia’s precise astrometry can identify candidate systems even when their individual components are not resolved in imaging. We therefore divide the candidates into two groups: resolved systems, in which the individual components have usable $r$-band measurements in DECaLS, PS1, NSC or SkyMapper, and unresolved systems, in which they are blended. We assign 376 systems to the resolved scoring route, using separately measured Legacy Surveys component photometry, and 321 to the unresolved route, using system-level photometry. The remaining 15 systems cannot be scored through either route owing to their data quality or available photometry. The 376 resolved and 321 unresolved systems are subsequently passed through the respective quasar colour analysis pipelines presented in Section~\ref{sec:p_colour}.

\subsection{Imaging}
Our second approach searches for multiplicity in the Tractor fitting models of DECaLS imaging. By searching for systems for which a point source cannot accurately fit the observed source, we remove the need for two resolved components and can expand our search to projected separations beyond the deblending limit of wide field surveys. 

For each quasar in the master catalogue, we searched for Tractor sources within $10^{\prime \prime}$ within objects with available DECaLS coverage. We selected systems containing at least two Tractor PSF components with usable positive $r$-band flux within a $1.5^{\prime \prime}$ radius.

The sources were initially ranked in descending $r$-band flux with the brightest source provisionally labelled as the primary object and the remaining component labelled as the secondary. The component labels were then revised to identify which component was more likely to be the quasar. At the redshift of each system, we calculated the median quasar $g-r$ colour using reference quasars within $\Delta z=\pm0.1$, requiring at least 20 reference objects.

For each component, its distance, $D_{g-r}$, from the expected $g-r$ quasar colour was calculated as

\begin{equation*}
D_{g-r,c}(z_0)=
\frac{|(g-r)_c-\widetilde{(g-r)}(z_0)|}{s_{g-r}(z_0)}.
\end{equation*}

where $(g-r)_{\mathrm{c}}$ and $\widetilde{(g-r)}(z_0)$ are the component and median quasar $g-r$ colours at redshift $z_0$, respectively. The reference quasar colour population is split into $\Delta z=\pm0.1$  and required at least 20 reference quasars. The colour scatter, $\sigma_{g-r}$, was defined using the 16th and 84th percentiles. The component with the smaller $D_{g-r}$ was defined as the final primary or quasar-like component and the other as the secondary.

After relabelling, we match the individual components to Gaia DR3 objects by crossmatching within $0.7^{\prime \prime}$ where available. The systems were subsequently divided into two-Gaia systems (where both components had different Gaia DR3 source IDs) and one-Gaia systems (where only the final primary had a Gaia match).

\subsubsection{Differential Chromatic Refraction}
An important consideration in dealing with multi-epoch ground based imaging such as DECaLS is the effect of Differential Chromatic Refraction (DCR). First raised as an issue for spectrophotometry by \citet{filippenkoImportanceAtmosphericDifferential1982}, DCR has become an important consideration for astronomical imaging. It appears as an atmospheric effect where the apparent position of an object shifts by a wavelength-dependent amount in the direction toward the zenith such that shorter wavelengths through a higher airmass are refracted slightly more than longer ones through lower airmass. 
The atmospheric refraction angle at wavelength, $\lambda$, for a given refractive index, $n(\lambda)$, with the true zenith distance, $z_t$ is approximated by

\begin{equation}
  R(\lambda)\:[\arcsec] \simeq 206265
  \left[
  \frac{n(\lambda)^2 - 1}{2n(\lambda)^2}
  \right]
  \tan z_t .
\end{equation}

The DCR can therefore be expressed as the difference between atmospheric refraction angles at wavelengths $\lambda_1$ and $\lambda_2$, given by

\begin{equation*}
\begin{aligned}
    \Delta R_{a,b}
    &= R(\lambda_1)-R(\lambda_2) \\
    &\simeq 206265
    \left[
        \frac{n(\lambda_1)^2 - 1}{2n(\lambda_1)^2}
        -
        \frac{n(\lambda_2)^2 - 1}{2n(\lambda_2)^2}
    \right] \tan z_t \:.
\end{aligned}
\end{equation*}

The DCR shift is along the zenith direction given by the parallactic angle such that DCR is stronger both at higher airmass (typically lower altitudes) and at shorter wavelengths. For each candidate system, we compute the proper angle between the two Tractor PSF components, $\theta_\mathrm{p}^\mathrm{T}$, and the estimated $\theta_\mathrm{p}^\mathrm{DCR}$ for the relevant $g$--band exposures. We therefore define the pair angle as the absolute difference between these angles,

\begin{equation*}
\Delta\theta_\mathrm{P} = \left |\theta_\mathrm{p}^\mathrm{T} - \theta_\mathrm{p}^\mathrm{DCR} \right | \:,
\end{equation*}

\begin{figure}
    \centering
    \includegraphics[width=1\linewidth]{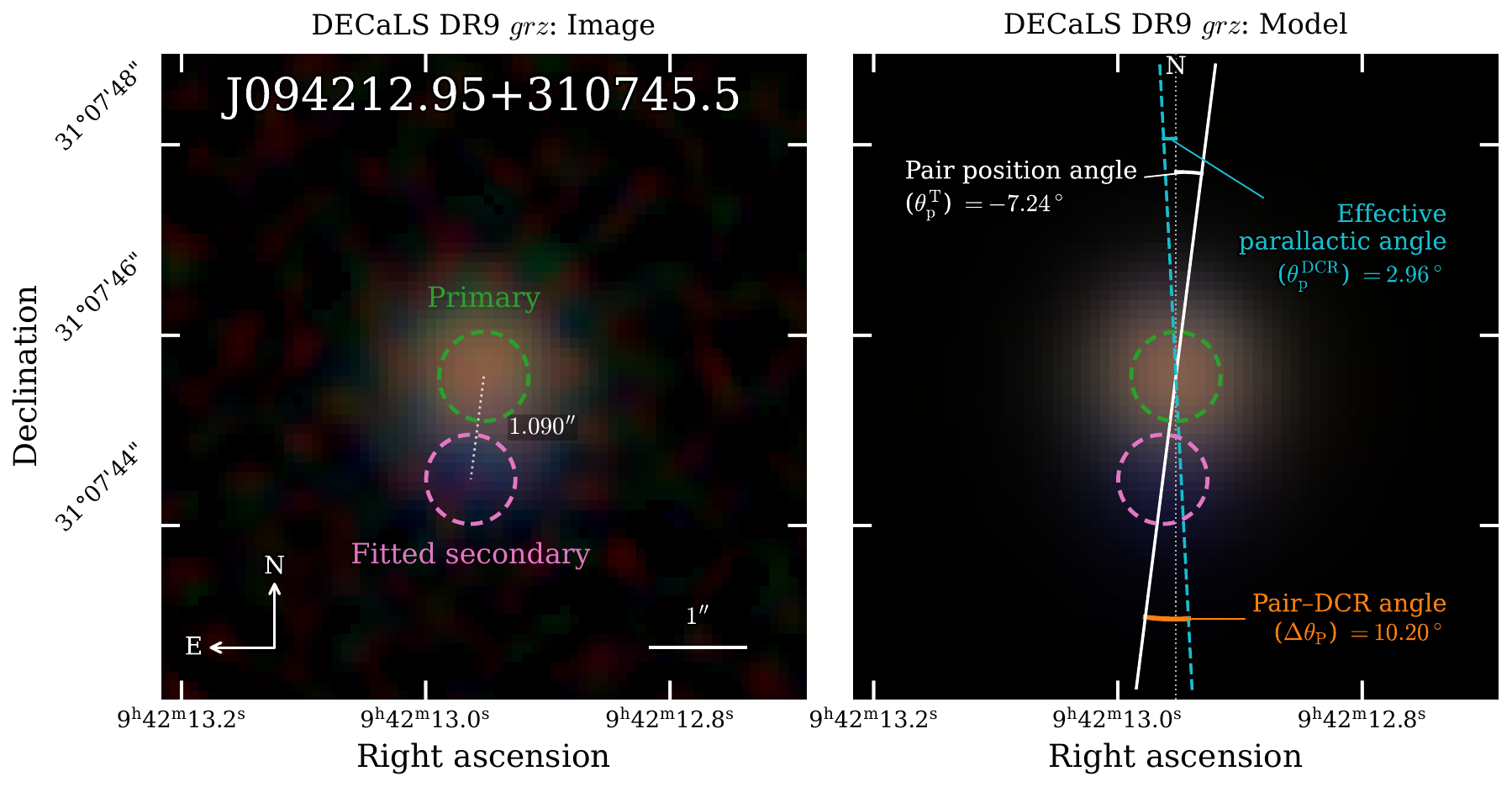}
    \caption{DECaLS DR9 $grz$ image (\textbf{left}) and corresponding Tractor model overlaid with DCR component axes (\textbf{right}) of suspected DCR contaminated object J094212.95+310745.5. Green and pink dashed circles mark the primary and spurious DCR secondary components respectively, separated by $1.090^{\prime\prime}$. In the model panel, the white solid line shows the pair axis, the cyan dashed line shows the DCR axis, and the white dotted line indicates North. The effective DCR direction is the average predicted direction of DCR distortion across the observations used to construct the coadd image. The orange arc marks the pair–DCR offset, $\Delta\theta_{\mathrm P}=10.20^\circ$, which fails the selection requirement $\Delta\theta_{\mathrm P}>20^\circ$. The blue extension and close alignment illustrate suspected DCR-related source contamination.}
    \label{fig:dcr_example}
\end{figure}

which ranges from $0\leq\theta^\circ\leq90$. The result of the DCR shift is that a single source can appear as two components with the spurious, bluer secondary component offset along the DCR direction as shown in Figure~\ref{fig:dcr_example}. For $\Delta\theta_\mathrm{P} \sim 0 ^\circ$, the two Tractor PSF components lie along the DCR direction suggesting that the observed multiplicity is likely due to DCR contamination. In such cases, a single quasar can appear stretched along the DCR angle in different colours and appear to have a companion. For $\Delta\theta_\mathrm{P} \sim 90 ^\circ$, the component pair axis is perpendicular to the DCR direction, thus suggesting that the observed multiplicity does not result from DCR contamination. This is important contamination guard when dealing with multi-epoch, ground based, coadd imaging data from DECaLS \citep{deyOverviewDESILegacy2019}. We therefore impose a minimum empirically defined pair angle threshold of $\Delta\theta_\mathrm{P}>20^\circ$ in our search which removes significant DCR contaminants while maintaining a significant population of candidates.

\subsubsection{Candidate selection}
Although the imposed pair angle cut, $\Delta\theta_\mathrm{P}\geq20^\circ$, demonstrated efficacy in removing DCR contaminants, the empirical nature of the threshold means that some remaining DCR contaminants are expected. To account for this, we impose a colour threshold based on nominal colour differences between the primary and secondary components in two-Gaia systems. We apply $|\Delta(g-r)_{\mathrm{s-p}}|\leq1$ as a preliminary colour cut to both one-Gaia and two-Gaia systems.

Owing to their extragalactic nature, we expect quasars to exhibit near-zero proper motion. Gaia's precision astrometry allows for extremely accurate proper motion measurements and we therefore apply a proper motion significance, $S_{\mathrm{PM}}$, threshold to our candidates. We quantify proper motion significance by comparing the measured motion in each sky coordinate with its uncertainty, while accounting for the correlation between the two measurements

\begin{equation}
    \label{eq:pmsig}
    S_{\mathrm{PM}}
    =
    \sqrt{
        \frac{
            \left(
                \dfrac{\mu_{\alpha\star}}
                {\sigma_{\mu_{\alpha\star}}}
            \right)^2
            +
            \left(
                \dfrac{\mu_{\delta}}
                {\sigma_{\mu_{\delta}}}
            \right)^2
            -
            2\rho
            \left(
                \dfrac{\mu_{\alpha\star} \mu_{\delta}}
                {\sigma_{\mu_{\alpha\star}}\sigma_{\mu_{\delta}}}
            \right)
        }{
            1-\rho^2
        }
    }.
\end{equation}

Here, $\mu_{\alpha\star}$ and $\mu_{\delta}$ are the proper motion
components in right ascension and declination,
$\sigma_{\mu_{\alpha\star}}$ and $\sigma_{\mu_{\delta}}$ are their
uncertainties, and $\rho$ is their correlation coefficient.
Larger values of $S_{\mathrm{PM}}$ indicate stronger evidence for
non-zero proper motion relative to the quoted uncertainties.

Components with redder colours in our quasar sample tend to have higher proper motion significances, as shown in Figure~\ref{fig:pmsig_distribution}, which is consistent with contamination from chance alignments with foreground stars. We therefore require that the companion component to known quasars in a candidate system must meet a proper motion significance of $S_{\mathrm{PM}_s}<3$ to reduce the contamination from foreground stars. We apply this cut to every system with two distinct Gaia sources: the resolved and blended one arcsecond routes and the two-Gaia imaging tier. Companions without a measured proper motion are kept and flagged (\NmissingPM final candidates). This threshold is empirically defined to reject contaminants and is discussed further in Section~\ref{sec:imaging_results}.

Even when unresolved in ground-based imaging, the Gaia-neighbour systems have separate Gaia detections. We therefore apply $S_{\mathrm{PM},s}<3$ to the secondary Gaia component where a finite measurement is available, retaining and flagging systems with missing secondary proper-motion significance.

\begin{figure}
    \centering
    \includegraphics[width=1\linewidth]{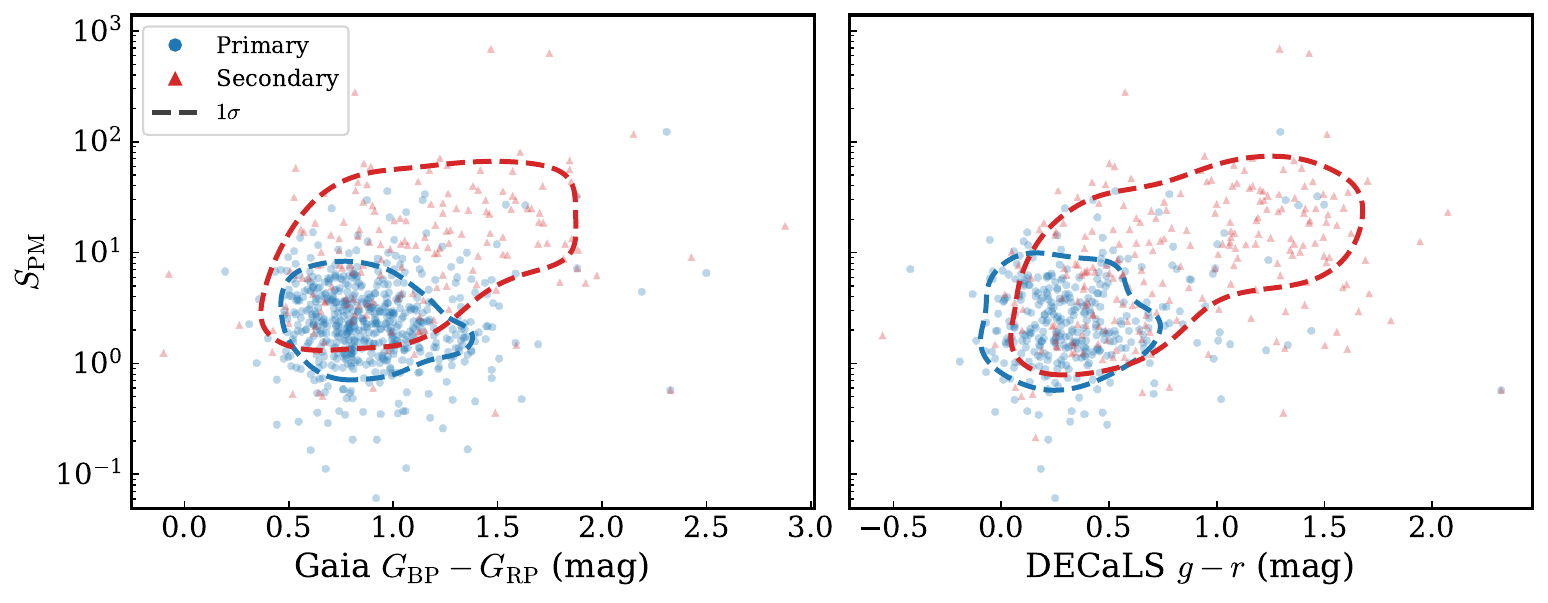}
    \caption{Gaia proper motion significance, $S_{\rm PM}$, as a function of colour from Gaia $G_{\rm BP}-G_{\rm RP}$ (\textbf{left}) and DECaLS $g-r$ (\textbf{right}) for primary (\textbf{blue}) and secondary (\textbf{red}) components in the one arcsecond input sample. The Gaia panel contains 574 primary and 163 secondary components and the DECaLS panel contains 343 primary and 225 secondary components. Dashed contours enclose $1\sigma$ of the distribution. Only components with usable proper motion and corresponding colour measurements are shown.}
    \label{fig:pmsig_distribution}
\end{figure}

\subsubsection{Morphological analysis}\label{sec:morphology}
ollowing the colour selection outlined in Section~\ref{sec:p_colour}, we examine the candidate morphologies since the smallest-separation systems are particularly susceptible to spurious source deblending artefacts. We therefore identify cases in which the fitted components may represent structure within a single extended source or contaminated PSF fitting rather than two distinct compact objects. We define two relative axes: the separation axis -- along which the two components are connected by the shortest separation -- and the axis perpendicular to this component separation axis, passing through the midpoint of the pair.

For each candidate, we construct two orthogonal, one-dimensional $r$-band profiles: one along the separation axis and another perpendicular through the pair midpoint. At each position, $x_i$, the profile flux, $f_i$, is the mean $r$-band pixel flux within a strip of $\pm1^{\prime \prime}$ around the corresponding axis. We estimate the local background level, $B$, from the outer 20\% at the extremities of each profile and define the non-negative flux weights as

\begin{equation*}
 w_i = \max(f_i-B,0).
\end{equation*}

The centre of the light distribution, $\bar{x}$, is calculated by giving greater weight to positions containing more flux, such that

\begin{equation*}
\bar{x}=\frac{\sum_i w_i x_i}{\sum_i w_i},\end{equation*}

The profile width, $\sigma$, then measures how widely the light is distributed around this centre.

\begin{equation*}
\sigma = \sqrt{\frac{\sum_i w_i(x_i-\bar{x})^2} {\sum_i w_i}}.
\end{equation*}

The profile width can be interpreted as the typical flux weighted distance of the light from its centre, with light farther from the centroid contributing more strongly. Finally, we calculate the ratio between the profile widths to classify the deviation from a single-source-like object and dual candidates,

\begin{equation*}
    R_{\perp}=\frac{\sigma_{\mathrm{along}}}{\sigma_{\perp}}.
\end{equation*}

For a single approximately symmetric source, the profile width measured along the proposed component-separation direction should be similar to the width measured perpendicular to it, giving $R_\perp \approx 1$. If a secondary source is present, its separation from the primary adds additional spread along the pair axis, increasing $\sigma_{\mathrm{along}}$. As a result, a clean dual system is expected to have $R_\perp>1$. While additional morphologies can produce $R_\perp>1$ (such as extended sources and projected alignments), the location of this analysis within the pipeline reduces the likelihood of contaminants from these sources.

\subsection{Quasar colour consistency}\label{sec:p_colour}
Quasar colours are known to occupy a relatively narrow distribution and have a well defined evolution across redshift, largely driven by their strong emission lines redshifting into or out of photometric bands as redshift changes. We use this intrinsic property of quasars to statistically quantify the photometric evidence for a second quasar consistent with the primary’s redshift. In resolved cases, we directly score the companion object relative to the nominal quasar colours at local redshift to determine whether that component is also a quasar at the parent's redshift, $z_0$, whereas a system--level analysis is required for blended, unresolved systems where we determine whether the blended photometry is better explained by two quasars or a quasar plus a foreground object. Contaminating spurious companions, such as stellar chance alignments, can alter the system's colours, allowing photometric model comparison to identify candidate quasar pairs. However, photometry alone cannot distinguish physical pairs from gravitational lenses. To distinguish intrinsically anomalous system colours from an apparent deviation caused by photometric measurement errors, one must deconvolve the measurement uncertainties from an observed reference population distribution to infer its intrinsic colour distribution, then account for each candidate’s own photometric uncertainties when assessing its consistency with that distribution.

We score colours with the \texttt{qso\_pcolor} package (\texttt{github.com/vasilybelokurov/qso\_p\_color}, revision \texttt{dbc3926}; later revisions change which sources are scored). We first build a reference population of 1,657,444 unique spectroscopically identified quasars from the position-deduplicated union of DESI DR1 and SDSS DR16Q and 2,978,857 cleaned point sources across 274 sky regions for the background. For increased model robustness, Quaia membership and infrared colours are used to remove additional likely quasars from the background training sample. The models are trained on 41 separate photometric bands from SDSS, Legacy Surveys, AllWISE, PS1, NSC, SkyMapper and VHS photometric surveys. Our candidates are scored on fewer bands. Resolved companions and imaging components are scored on the Legacy Surveys $g$, $r$ and $z$ bands only, because the other surveys do not resolve sub-arcsecond pairs. Blended systems are scored on Legacy Surveys photometry together with PS1, NSC and SkyMapper total-light magnitudes (Section~\ref{sec:unresolved-method}). Photometric measurements are retained when they have finite fluxes and valid positive uncertainties, and satisfy the relevant survey specific quality requirements. Invalid measurements are treated as missing, and their bands are excluded from the likelihood. Eligibility requires reference $r-$band S/N ratios of $\geq10$ and $\geq20$ for resolved and unresolved scoring respectively while usable measurements in other bands are retained without an additional S/N requirement.

Before modelling, we correct the catalogue photometry for Galactic
extinction using the SFD98 reddening map \citep{schlegelMapsDustInfrared1998} and band specific extinction
coefficients. For band $j$, the corrected flux and variance are
$f_j^0=f_j10^{0.4R_jE(B-V)}$ and
$(\sigma_{f,j}^0)^2=\sigma_{f,j}^2 10^{0.8R_jE(B-V)}$,
respectively. We then transform the corrected fluxes into luptitudes,

\begin{equation*}
\ell_j =
22.5-\frac{2.5}{\ln 10}
\left[
\operatorname{asinh}\left(\frac{f_j^0}{2s_j}\right)
+\ln s_j
\right],
\end{equation*}

where $s_j$ is the softening parameter in band $j$. Luptitudes (denoted $\ell_x$ where $x$ represents the photometric band) offer a useful alternative to magnitude analysis for our use case owing to their property of approaching conventional magnitudes at high flux but remaining finite for zero or negative fluxes, allowing weak measurements to contribute without imposing a positive flux cut.

We represent the intrinsic population distributions of quasars in colour--redshift space in 43 overlapping redshift slices covering $0.15\leq z_k\leq4.35$ with centres separated by $\Delta z=0.1$ using Gaussian mixtures. Each of the 43 slices includes quasars within $z_k=\pm0.1$ of its centre, and for each slice we fit a Gaussian mixture with $K$ components, ranging from 4 to 20 depending on the number of objects within the slice, to describe correlations between bands and different amounts of scatter:

\begin{equation*}
p(\boldsymbol{x}\mid Q,z_k)
=
\sum_{i=1}^{K}
\alpha_i\,
\mathcal{N}\!\left(
    \boldsymbol{x};
    \boldsymbol{\mu}_i,V_i
\right),
\end{equation*}

However, it may be that the distribution of objects is impacted by catalogue measurement errors rather than an intrinsic population distribution spread. To account for this, we deconvolve the intrinsic photometric distribution from the measurement errors using extreme deconvolution \citep[XD, ][]{bovyExtremeDeconvolutionInferring2011} to estimate the underlying distribution while accounting for each object’s measurement covariance and available bands. For object $n$, the observed data likelihood is

\begin{equation*}
p(\boldsymbol{y}_n\mid Q,z_k)
=
\sum_{i=1}^{K}
\alpha_i\,
\mathcal{N}\!\left(
    \boldsymbol{y}_n;
    H_n\boldsymbol{\mu}_i+\boldsymbol{b}_n,\,
    H_nV_iH_n^{\mathsf T}+T_n+C_n
\right),
\end{equation*}

where $H_n$ maps the latent coordinates into the object’s observed native bands, $\boldsymbol b_n$ and $T_n$ describe calibration offsets and instrumental scatter and $C_n$ is the measurement covariance matrix in luptitude space. This allows our model to distinguish intrinsic population scatter from noise induced scatter.

We developed two complementary quasar models, one with and one without magnitude dependence. The models are identical in bands, extinction, background, priors and faint limits but differ in their assumption on whether a quasar's colour at fixed redshift is dependent on its apparent magnitude. The magnitude dependent model assumes the intrinsic quasar photometry is represented by a joint Gaussian mixture of magnitude and colours within each redshift slice such that the predicted colour distribution may vary with apparent magnitude. While this captures genuine brightness dependent behaviour, it may be susceptible to spurious magnitude dependent selection effects. The magnitude independent model constrains the mixture such that the colour distribution at fixed redshift is independent of magnitude.

We also developed a "background model" to represent the photometric distribution of field point sources, predominantly stars, that can contaminate the companion sample. The background reference population comprises 2,978,857 cleaned point sources, of which 1,975,894 are assigned to fitting and model selection while the remainder are reserved for calibration and testing. Known and likely quasars are removed to prevent the model from learning quasar colours as ordinary background behaviour but it still may contain unidentified ones, especially at fainter magnitudes. Because faint sources dominate the catalogue and the stellar colour distribution changes with brightness, the background model is fitted separately in eight DECaLS $r$ bins spanning $15.5<r<24.5$ with 20 Gaussian components per bin applied at $|b|\geq25^\circ$. These are combined into 

\begin{equation*}
p_B(\boldsymbol{x}\mid l,b)=\sum_{i=1}^{160}w_i(l,b)\,\mathcal{N}(\boldsymbol{x};\boldsymbol{\mu}_i,V_i),
\end{equation*}

where $\boldsymbol{x}$ represents the source’s intrinsic brightness across 36 modelled photometric bands, before measurement noise is added. The 41 observed channels become the 36 latent coordinates because the ten DECaLS North/South channels map to five shared bands. XD accounts for measurement errors and missing bands during fitting.

\subsubsection{Resolved systems}\label{sec:resolved-method}
For systems for which both components resolved such that independent photometry may be measured for each object, we developed a probabilistic method to assess whether the companion is consistent with a quasar at the primary's adopted redshift, $z_0$. Eligible companion components must have point source morphology, $|b|\geq25^\circ$, and reference band S/N$\geq10$ and define consistency with the primary's redshift using a rest-frame velocity window of $\pm2000\,\mathrm{km\,s^{-1}}$ following the threshold defined in \citet{hennawiBinaryQuasarsSloan2006}, corresponding to a redshift half-width

\begin{equation*}
\delta z = \frac{(1+z_0)\,\Delta v}{c},
\qquad
\Delta v=2000\,\mathrm{km\,s^{-1}}.
\end{equation*}

For primaries with photometric Quaia redshifts, the redshift error (median of $\sigma_z\approx0.04(1+z)$ for the final selected Quaia sample) is larger than this window. For resolved components we add the published per-object Quaia error in quadrature to the half-width of the window. For blended systems we average the QQ and QS likelihoods separately over nine redshift nodes within $\pm3\sigma_z$, with Gaussian weights. Quaia candidates therefore have a weaker redshift constraint, and we report them separately.

At the companion’s measured $r$-band brightness, we evaluate the joint likelihood of its measurements in the other available bands by conditioning each population model on the reference magnitude, $m_r$. Denoting these remaining measurements by $D$ and the quasar or background hypothesis by $H$, the conditional likelihood is

\begin{equation*}
p(D\mid H,m_r)=\frac{p(D,m_r\mid H)}{p(m_r\mid H)}.
\end{equation*}

We compare four explanations for the companion: a quasar near $z_0$, a quasar at another redshift, a background point source, and a broad outlier population. Each explanation is assigned an intensity combining its photometric likelihood with its population abundance. For example, the quasar intensity is

\begin{equation*}
\lambda_Q(z)
=
\Sigma_Q(z,m_r)\,
p(D\mid Q,z,m_r),
\end{equation*}

where $\Sigma_Q(z,m_r)$ is the expected quasar surface density per unit redshift and reference magnitude. We take $\Sigma_Q$ from the eBOSS quasar luminosity function (PLE+LEDE fit of \citealt{palanque-delabrouilleExtendedBaryonOscillation2016}), and count the background surface density directly from the Legacy Surveys catalogue at the object's position. The companion ranking statistic is

\begin{equation*}
\mathcal R=
\frac{\langle\lambda_Q\rangle_{z_0\pm\delta z}}
{\displaystyle\int\lambda_Q(z)\,dz+\lambda_B+\lambda_U}.
\end{equation*}

where $\langle\lambda_Q\rangle_{z_0\pm\delta z}$ is the average quasar intensity over the redshift window, $z_0\pm\delta z$, and $\lambda_B$ and $\lambda_U$ are the background and outlier intensities. Thus, the denominator includes quasars at all modelled redshifts as well as non-quasar explanations. Two tests are applied before ranking. First, a companion whose colours lie outside the region occupied by quasars at $z_0$ is not scored: we draw 255 simulated quasars at $z_0$ in the companion's bands, add its measurement errors, and reject the companion if its lower-tail rank fraction is below $2/256=0.0078125$, equivalent to the 0.78125th percentile. Second, a companion that resembles neither population is not scored: for quasars at any redshift and for the background, we compare its distance from the nearest mixture component with the same distance for 256 simulated members observed in its bands, and reject it when both tail fractions are below $2/256$. We rank companions using $\ln\mathcal{R}$. The corresponding model probability for the finite redshift window is $P_{\mathrm{same}}=\mathcal{R}\,\Delta z_{\mathrm{eff}}$, where $\Delta z_{\mathrm{eff}}$ is the effective window width.

We define the calibrated scoring statistic as the maximum finite value of $\ln\mathcal R$ in the available sample of candidates with Gaia astrometry under either the magnitude dependent or independent models. We hypothesise that these remaining candidates may be grouped into two classifications: a quasar pair or a quasar with a projected contaminant. We therefore fit two Gaussians to the distribution of system's $\ln\mathcal{R}$ score as a function of the secondary component's proper motion significance. Figure~\ref{fig:ln_r_vs_pm_sig} shows the $1\sigma$ ellipses of the two fitted Gaussians, which represent the two expected classes of objects. We find that there are broadly two distinct regions of the distribution: A region with high $\ln\mathcal{R}$ at lower secondary $S_{PM}$ and a broader region of systems with lower $\ln\mathcal{R}$ and higher secondary $S_{PM}$. Given that quasars should demonstrate near-zero $S_{PM}$, we deduce that the large tail of systems at high secondary $S_{PM}$ are likely contaminated with spurious secondary components.

\begin{figure}
    \centering
    \includegraphics[width=1\linewidth]{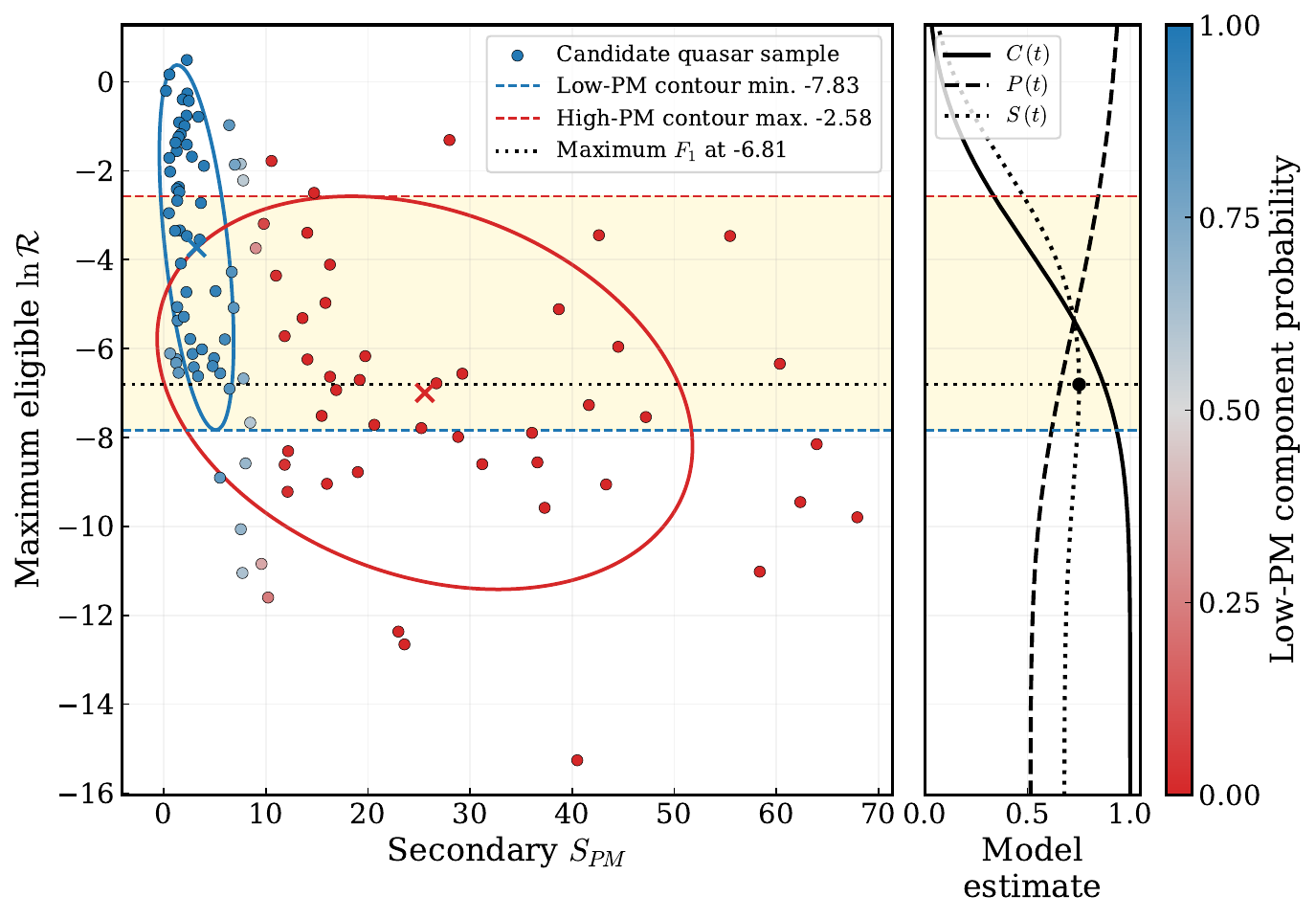}
    \caption{Calibration of the resolved companion score, $\ln\mathcal{R}$ (the larger of the two models' values). \textbf{Left:} scores versus secondary proper motion significance, $S_{PM}$ for \CalibN one arcsecond systems with usable $S_{PM}$ measurements with colours indicating the probability of belonging to the low-$S_{PM}$ population from red (0) to blue (1). The blue and red ellipses enclose $1\sigma$ of the fitted low- and high-$S_{PM}$ Gaussian components, respectively. The minimum score on the blue ellipse and maximum score on the red ellipse define the completeness, $C(t)$, and purity, $P(t)$, thresholds and thus the search interval of \CalibInterval. \textbf{Right:} Gaussian-based estimates of $C(t)$ (solid), $P(t)$ (dashed), and their harmonic mean $S(t)$ (dotted), calculated by integrating the fitted score distributions above each trial cut $\ln\mathcal{R}\geq t$. The low-$S_{PM}$ component serves as the quasar-like population and the high-$S_{PM}$ component as the contaminant-like population. The black dotted horizontal line marks the adopted threshold, $\ln\mathcal{R}=$\LnRthreshold.}
    \label{fig:ln_r_vs_pm_sig}
\end{figure}

The calibration for $\ln\mathcal{R}$ upon which we define a threshold for candidates may be selected following two approaches. To quantify the trade-off between completeness and purity as a function of the selection threshold, we use the two fitted $1\sigma$ ellipses as representatives for the quasar (blue) and contaminant (red) populations. The completeness-driven approach uses the lower limit of the $1\sigma$ quasar population ellipse while the purity approach cuts at the maximum value of the contaminant population ellipse in $\ln\mathcal R$. For each trial cut $\ln\mathcal{R}\geq t$, we calculate the fraction of objects remaining from the completeness sample, $C(t)$. From this we calculate the purity, $P(t)$, as the number of retained quasar population objects, $N_{qso}$, divided by the total retained from both $N_{qso}$ and the contaminant population, $N_{cont}$, accounting for their fitted population weights, expressed in equation.

\begin{equation*}
C(t)=\frac{N_{qso}(t)}{N_{qso}(-\infty)},
\qquad
P(t)=\frac{N_{qso}(t)}{N_{qso}(t)+N_{cont}(t)}.
\end{equation*}

Repeating this calculation across the $\ln\mathcal R$ range produces the completeness and purity curves. We combine the purity and completeness into one score as

\begin{equation*}
S(t)=\frac{2P(t)C(t)}{P(t)+C(t)}.
\end{equation*}

We select a threshold of $\ln\mathcal{R}=$\LnRthreshold, which maximises $S$ with model-estimated completeness of \CalibCompleteness and purity of \CalibPurity, fitted to \CalibN systems. This threshold is subsequently applied to all resolved component colour analysis scoring including components for which astrometry is not available.

\subsubsection{Unresolved systems}\label{sec:unresolved-method}
Different selection methodologies and criteria may produce candidates which are unresolved in current photometric surveys. We therefore develop a complementary scoring system to evaluate the candidacy of such systems where individual components are unresolved but system level photometry is available. The underlying assumption with this method is that the measured catalogue photometry captures both components. However, high-resolution follow-up imaging would be required for a more robust downsampling of candidates. We use DECaLS photometry supplemented with PS1, NSC and SkyMapper. Each band must measure the total light of both components, so we use the Tractor model flux in DECaLS, Kron magnitudes in PS1, \texttt{MAG\_AUTO} in NSC and Petrosian magnitudes in SkyMapper. The models were fitted to PSF photometry of single sources; for blends smaller than the seeing the total flux should equal the sum of the components' PSF fluxes, but this has not been tested on real blends. Without component photometry, precise groupings become difficult so we compare two hypotheses. Under $H_{QQ}$, the system is built of a quasar pair at system redshift $z_0$ while under $H_{QS}$, it comprises a known quasar and a companion drawn from the stellar dominated background population. 

Spectroscopic redshift measurements are unavailable for the unresolved companion object. Therefore under $H_{\mathrm{QQ}}$, both components are assumed to lie at the adopted primary redshift $z_0$ and quasar pairs at different redshifts are not included as a separate hypothesis. We require both model components to contribute at least ten times the uncertainty in the total reference band flux, which requires a combined $S/N\geq20$. We remove the requirement for point source morphology as the blended system may appear elongated along the axis of separation. Additionally, to reduce the impact of increased contamination from the dense distribution of stars on the galactic plane, we also require that $|b|\geq25^\circ$.

At this point, we are still unsure of the contribution of flux from the secondary component, so we introduce a flexible companion flux share parameter, $\alpha$, denoting the fractional contribution of the reference flux from the secondary,

\begin{equation*}
\frac{10\sigma_t}{F_t}\leq\alpha\leq1-\frac{10\sigma_t}{F_t},
\end{equation*}

where $\sigma_t$ is the uncertainty in the total flux, $F_t$, in the reference band (e.g. Legacy Imaging Surveys' $r$-band).

As such, the fluxes of the primary and secondary components may be expressed as

\begin{equation*}
F_1=(1-\alpha)F_t,\qquad F_2=\alpha F_t,
\end{equation*}

At each value of $\alpha$, the assumed division of the total reference band flux determines the individual component brightnesses. We use these brightnesses to predict the combined photometry under both hypotheses: $H_{\mathrm{QQ}}$, using the quasar model at $z_0$ for both components, and $H_{\mathrm{QS}}$, using the quasar model and the background model for the primary and secondary objects respectively. This calculation is repeated using both the magnitude dependent and magnitude independent models.

For each trial reference band flux split, we predict and sum the component fluxes in the other available bands and evaluate their agreement with the observed combined photometry, accounting for intrinsic photometric scatter and measurement uncertainties. The combined photometric likelihood is evaluated in luptitude space using a first order approximation to flux addition, including intrinsic photometric scatter and measurement uncertainties.

We weight each trial flux split using the population brightness priors. Under $H_{\mathrm{QQ}}$, both components follow the quasar brightness distribution at $z_0$ whereas under $H_{\mathrm{QS}}$, the companion follows the background brightness distribution. The companion prior is normalised over its detectable brightness range.

We numerically integrate the weighted likelihoods over the allowed $\alpha$ values on 48 points evenly spaced in $\ln[\alpha/(1-\alpha)]$, including the required coordinate-conversion factors. This gives separate evidence values, $L_{\mathrm{QQ}}$ and $L_{\mathrm{QS}}$. For systems without Legacy $r$, we choose the first available fallback band in the order PS1 $r$, NSC $r$ then SkyMapper $r$, all requiring $S/N\geq20$. In such cases, we treat Legacy $r$ as unknown and integrate over 41 trial luptitudes spanning $\pm(1.2\,\mathrm{mag}+5\sigma_{\ell,\mathrm{fb}})$ around the fallback luptitude, where $\sigma_{\ell,\mathrm{fb}}$ is the fallback measurement’s uncertainty in luptitudes.

We use the Bayes factor, in Equation~\ref{eq:bayes_qq_qs}, to compare the evidence for $H_{\mathrm{QQ}}$ and $H_{\mathrm{QS}}$, with positive $\ln\mathrm{BF}_{\mathrm{QQ,QS}}$ favouring a quasar pair at the same redshift and negative values favouring a quasar with a background companion.

\begin{equation}\label{eq:bayes_qq_qs}
\ln\mathrm{BF}_{\mathrm{QQ,QS}}
=\ln\left(\frac{L_{\mathrm{QQ}}}{L_{\mathrm{QS}}}\right).
\end{equation}

Selecting $\ln\mathrm{BF}_{\mathrm{QQ,QS}}>0$ corresponds to equal prior odds for a quasar and a star companion. We retain systems with finite Bayes factors from both model variants for which $H_{\mathrm{QQ}}$ is favoured by at least one variant while noting that systems favouring $H_{\mathrm{QQ}}$ in both models are preferred. It's also worth noting that $QQ$ preference prioritises follow-up observation and analysis but does not establish a physical dual quasar by itself. High resolution follow-up imaging is needed to assess whether the catalogue measurements capture both components.

\section{Results and Discussions}
\subsection{Results of the one arcsecond Gaia search}
The Gaia one arcsecond neighbour search contains 712 initial candidate systems. We evaluated these using the photometric population models outlined in Section~\ref{sec:p_colour}, assigning 376 entries to the resolved and 321 to the unresolved photometry pathways respectively. The remaining 15 had insufficient component photometry for the resolved pathway and were not scored through either adopted route.

\subsubsection{Resolved systems}
After applying the photometric, galactic latitude and morphology cuts, we obtain eligible scores for \NresolvedEligibleMagdep and \NresolvedEligibleMagindep systems from the magnitude-dependent and magnitude-independent models, respectively. Applying $\ln\mathcal R\geq$\LnRthreshold, calculated in Section~\ref{sec:resolved-method}, retains \NresolvedScoreMagdep and \NresolvedScoreMagindep systems. Following the score selection, we require the secondary Gaia proper-motion significance to satisfy $S_{\mathrm{PM},s}<3$ where a finite measurement is available, retaining and flagging systems with missing measurements. This leaves 94 systems selected by the magnitude-dependent model and 95 by the magnitude-independent model, with \NpreResolved passing at least one variant. We adopt this either-model selection. Removing
confirmed non-lens literature contaminants leaves \Nresolved resolved candidates in the final catalogue; confirmed lenses are retained and flagged.

\begin{table}
    \centering
    \small
    \renewcommand{\arraystretch}{1.25}
    \caption{Cumulative candidate counts through the resolved one arcsecond pipeline. "Either model" counts unique systems passing at least one model. The same numerical threshold, $\ln\mathcal{R}\geq$\LnRthreshold, is applied to both model variants. The secondary Gaia proper-motion cut is applied after the score threshold; systems with missing measurements are retained and flagged.}
    \label{tab:one_arcsec_resolved_cutflow}
    \begin{tabularx}{\linewidth}{@{}Xrrr@{}}
        \toprule
        \textbf{Selection stage}
            & \shortstack{\textbf{Magnitude}\\\textbf{dependent}}
            & \shortstack{\textbf{Magnitude}\\\textbf{independent}}
            & \shortstack{\textbf{Either}\\\textbf{model}} \\
        \midrule
        $S/N\geq10$
            & 376 & 376 & 376 \\
        Galactic latitude $|b|\geq25^\circ$
            & 309 & 309 & 309 \\
        Companion PSF morphology
            & 288 & 288 & 288 \\
        Remaining model eligibility and support requirements
            & 172 & 190 & 195 \\
        $\ln\mathcal{R}\geq$\LnRthreshold
            & 123 & 135 & 144 \\
        Secondary Gaia $S_{\mathrm{PM},s}<3$; missing retained
            & 94 & 95 & 101 \\
        After literature exclusions
            & 91 & 92 & \textbf{98} \\
        \bottomrule
    \end{tabularx}
\end{table}

The resolved candidates span a redshift range of \zRangeRes (which is largely limited by the initial redshift selection criteria) with a median redshift of \zMedRes. The separation range of the candidate systems spans \SepRangeRes$^{\prime\prime}$, corresponding to projected physical distances of \kpcRangeRes kpc assuming the Planck18 cosmology, with $H_0=67.66\mathrm{km s^{-1} Mpc^{-1}}$ and $\Omega_m=0.30966$.

\begin{figure}
    \centering
    \includegraphics[width=1\linewidth]{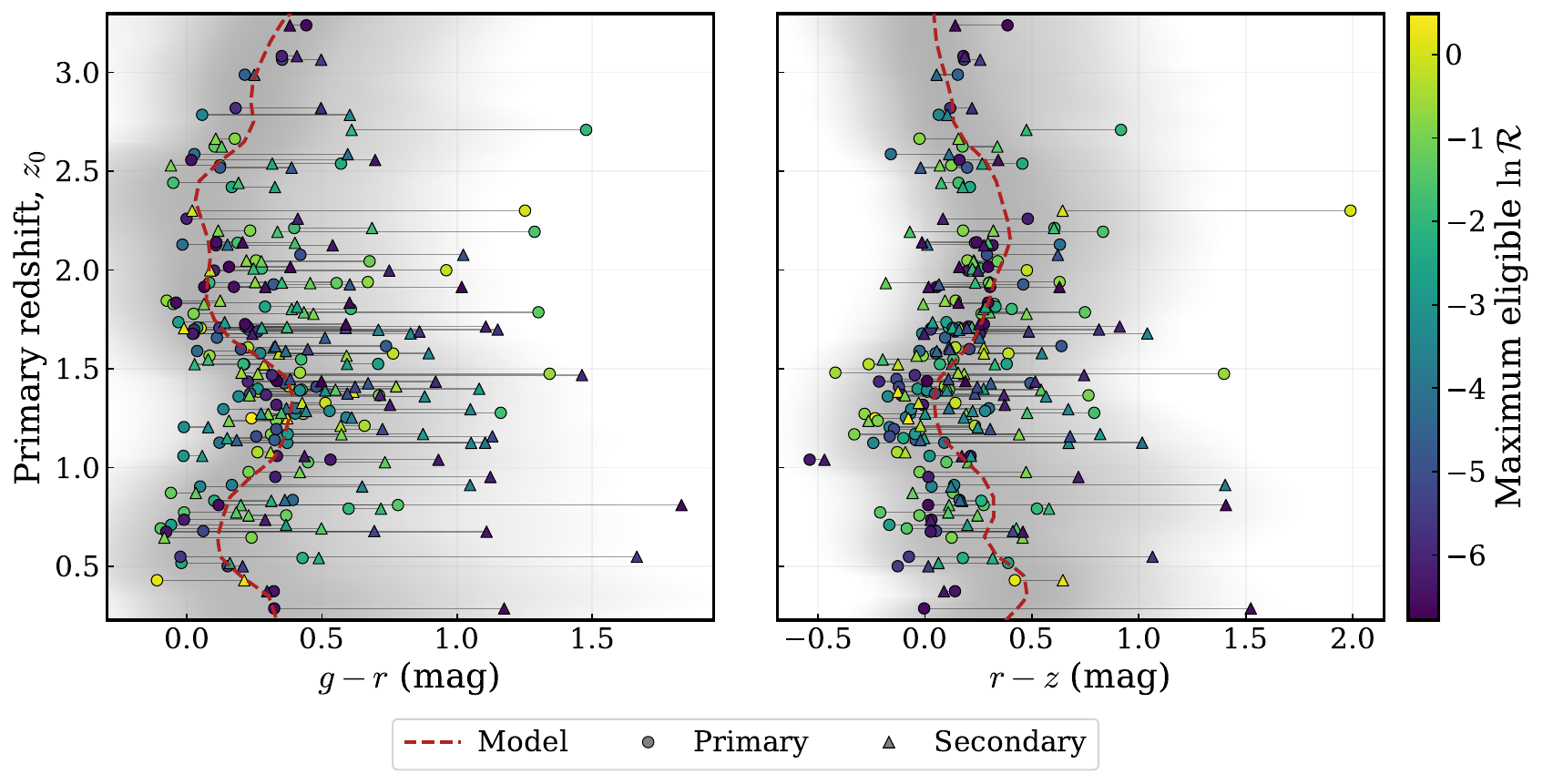}
    \caption{DECaLS $g-r$ and $r-z$ colours as a function of the adopted primary redshift for \NresolvedAboveThreshold systems in the resolved subset satisfying $\ln\mathcal R\geq-6.81$. Circles and triangles denote primary and secondary components respectively, coloured by the larger eligible companion $\ln\mathcal R$ from the magnitude-dependent and magnitude-independent models and with components belonging to the same system connected by black lines. Grey shading illustrates the quasar-model colour distribution, and the dashed red curve shows the median of the DECaLS North model.}
    \label{fig:colour_redshift_one_arcsec}
\end{figure}

\subsubsection{Unresolved systems}
Of the 321 entries assigned to the unresolved pathway, \NblendedFinite yield finite QQ--QS Bayes factors from both model variants. The magnitude-dependent and magnitude-independent models favour QQ for \NblendedQQMagdep and \NblendedQQMagindep systems, respectively, with \NblendedQQBoth favoured by both and \NblendedQQEither favoured by at least one. We adopt the either-model selection. Applying the secondary Gaia proper-motion requirement, while retaining and flagging systems with missing measurements, leaves \NpreBlended candidates. Following the literature contaminant exclusions and the removal of \Nartefact corrupted-data artefact identified during visual inspection, the final unresolved sample contains \Nblended candidates. Among these final candidates, \NblendedLegacyR use Legacy-$r$ photometry directly, while \NblendedFallback require fallback photometry: \NblendedPSone use PS1-$r$ and \NblendedNSC use NSC-$r$.

\begin{table}
    \centering
    \small
    \renewcommand{\arraystretch}{1.25}
    \caption{Cumulative candidate counts through the unresolved one-arcsecond pipeline. The final selection requires finite Bayes factors from both model variants and “either model” counts systems for which at least one gives $\ln B_{\mathrm{QQ,QS}}>0$.}
    \label{tab:one_arcsec_unresolved_cutflow}
    \begin{tabularx}{\linewidth}{@{}Xrrr@{}}
        \toprule
        \textbf{Selection stage}
            & \shortstack{\textbf{Magnitude}\\\textbf{dependent}}
            & \shortstack{\textbf{Magnitude}\\\textbf{independent}}
            & \shortstack{\textbf{Either}\\\textbf{model}} \\
        \midrule
        Assigned to unresolved route
            & 321 & 321 & 321 \\
        Galactic latitude $|b|\geq25^\circ$
            & 149 & 149 & 149 \\
        Usable reference band photometry
            & 147 & 147 & 147 \\
        Redshift within model support
            & 144 & 144 & 144 \\
        $S/N\geq20$
            & 142 & 142 & 142 \\
        Finite $\ln B_{\mathrm{QQ,QS}}$ in both variants
            & 140 & 140 & 140 \\
        QQ favoured ($\ln B_{\mathrm{QQ,QS}}>0$)
            & 36 & 36 & 45 \\
        Secondary Gaia $S_{\mathrm{PM},s}<3$; missing retained
            & 32 & 31 & 40 \\
        After literature exclusions
            & 31 & 30 & 39 \\
        After corrupted-data artefact removal
            & 30 & 30 & \textbf{38} \\
        \bottomrule
    \end{tabularx}
\end{table}

\subsection{Results of the imaging search}\label{sec:imaging_results}
The prevalence of the DCR contamination is apparent in the 'blue tails' in  $\Delta (g-r)$ as a function of redshift in Figure~\ref{fig:channel1a_delta_g_minus_r_vs_redshift_and_delta_theta} as well as in a large clump of systems with component separations in the range $0.6 \lesssim \Delta\theta \lesssim 1.3$. Spurious blue secondary components resulting from the DCR shift in multi-epoch coadd imaging push these systems to low $\Delta (g-r)$. The blue tails occur at the redshifts where strong emission lines move through the $g$ band. This is expected: an emission line in the $g$ band shifts the effective wavelength of a quasar's $g$-band light and so changes its refraction relative to a star \citep{kaczmarczikAstrometricRedshiftsQuasars2009}. We apply the DCR cut (angle $>20^\circ$) once, and only to one-Gaia imaging systems; two-Gaia systems are exempt because Gaia observes from space.

\begin{figure}
    \centering
    \includegraphics[width=1\linewidth]{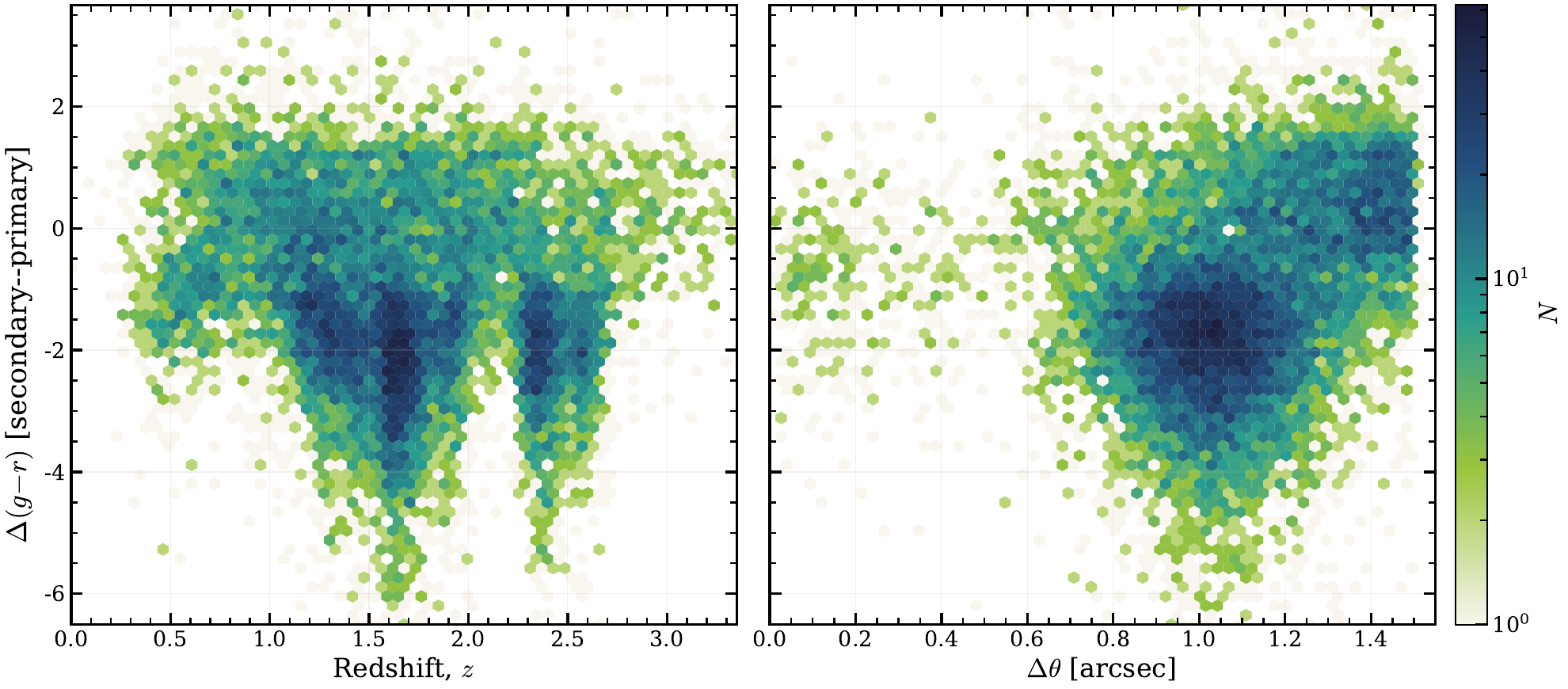}
    \caption{Density of $\Delta(g\!-\!r)$ as a function of (\textbf{left}) vs $z$ for 13,803 dual quasar candidate systems with two DECaLS Tractor PSF components. \textbf{(right)} component separation, $\Delta\theta$. The long blue tails at $0.8\lesssim z \lesssim 2$ and $z \approx 2.5$ with the associated clump of systems that have $0.6 \lesssim \Delta\theta \lesssim 1.3$ at $\Delta (g-r)\lesssim-1$ appear to be blue secondary components to the primary quasar which we find to be spurious contamination from DCR effects.}
    \label{fig:channel1a_delta_g_minus_r_vs_redshift_and_delta_theta}
\end{figure}

However, rejecting systems with $\Delta\theta_{\mathrm P}\leq20^\circ$, thereby retaining $\Delta\theta_{\mathrm P}>20^\circ$, removes much of this contamination -- see Figure~\ref{fig:3panel}. Despite this, some systems with large, negative $\Delta (g-r)_{s-p}$ are still present. When analysing candidate systems for which both components have Gaia IDs, we find that most exhibit a narrow range of component colour difference with $|\Delta (g-r)_{s-p}|\leq2$ (outlined in green in Figure~\ref{fig:3panel}). However, we hypothesise that companion secondary quasars at a similar redshift to the primary will exhibit near identical colours to those of the primary. We therefore apply a threshold cut at $|\Delta (g-r)|\leq1$ both to remove the remaining DCR contamination and improve the purity of the final candidate sample. 

\begin{figure*}
    \centering
    \includegraphics[width=\linewidth]{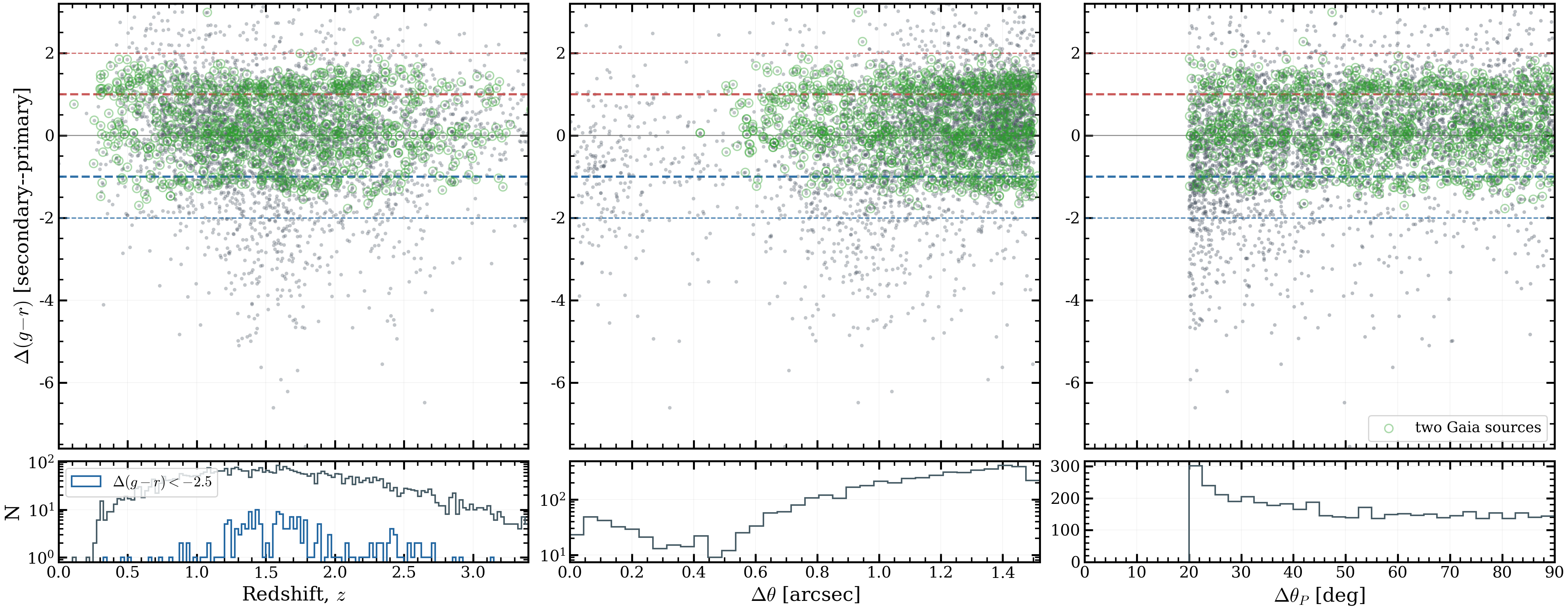}
    \caption{Distribution of component colour differences for DECaLS Tractor dual PSF quasar systems as a function of redshift, component separation, $\Delta\theta$, and DCR alignment, $\Delta\theta_\mathrm{P}$, after applying the $\Delta\theta_\mathrm{P}>20^\circ$ cut. The highlighted Gaia subsample in green marks systems in which both Tractor PSF components have distinct Gaia counterparts.}
    \label{fig:3panel}
\end{figure*}

We apply further constraints on our sample using proper motion significance based on the hypothesis that quasars should exhibit zero or near-zero intrinsic proper motion at the sub-mas precision. As previously discussed, however, a measured non-zero proper motion does not necessarily indicate genuine motion because it may arise from measurement uncertainty. We therefore apply proper motion significance, which compares the proper motion measurement with its uncertainty and thus corrects for measurement errors.

We find that, in the two-Gaia sample, $\Delta(g-r)_{s-p}$ systematically increases for secondary components at higher proper motion significances, consistent with contamination from foreground stars which commonly exhibit redder colours and statistically significant proper motions, entering the sample as projected secondary components. We require that $S_\mathrm{PM}<3$ and $|\Delta(g-r)_{s-p}|\leq1$. These cuts empirically retain systems whose components have statistically insignificant proper motions and broadly consistent colours, while rejecting red, significantly moving companions that are more likely to be foreground stars.

\begin{figure}
    \centering
    \includegraphics[width=1\linewidth]{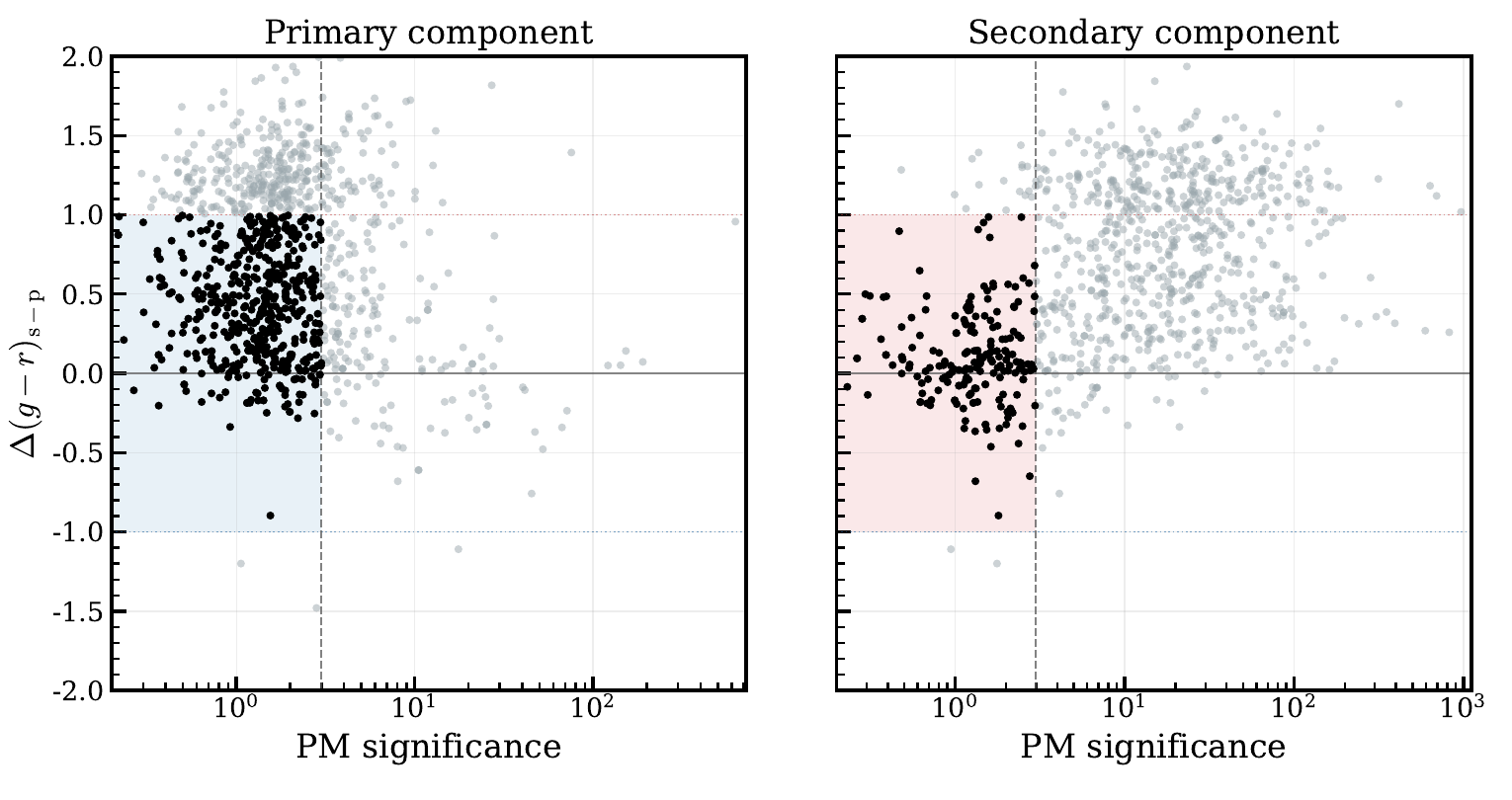}
    \caption{Gaia proper motion significance against the component colour difference, $\Delta(g-r)_{\mathrm{s-p}}$, for systems with two distinct Gaia DR3 detections. The left and right panels show the proper motion significance of the primary and secondary components respectively. The blue and red shaded regions highlight the candidate selection regions that satisfy $|\Delta(g-r)_{\mathrm{s-p}}|\leq1$ and  $S_{\mathrm{PM}}<3$. The highlighted subset contains 491 primary and 210 secondary components.}
    \label{fig:pmsig_primary_secondary}
\end{figure}

The above pipeline takes the 14,369 initial systems with two Tractor PSF components and downselects to \NimagingPreDedupOne one-Gaia and \NimagingPreDedupTwo two-Gaia systems. In some systems with wide component separations, both the primary and secondary components may be independently detected by the parent catalogues, leading to two duplicate system entries in the final candidate list. We therefore deduplicate such rows according to the smallest separation between the selected primary component and the parent-catalogue position (falling back to the smallest $|\Delta(g-r)_{s-p}|$, if the separations are tied).

We evaluate the \NimagingScored systems surviving the preliminary astrometric, colour-difference and tier-specific DCR cuts using the resolved-source models in \texttt{qso\_pcolor}, described in Section~\ref{sec:p_colour}. We retain the primary and secondary assignments established earlier in the pipeline and recover each component's separately fitted Legacy Surveys $g,r,z$ fluxes and uncertainties using its saved Tractor identifiers. Measurements retain their native North or South photometric system. The package applies Galactic-extinction correction, transforms the fluxes to luptitudes and includes the measurement uncertainties when evaluating the population models.

Both components must satisfy the reference $r$-band $S/N\geq10$, Galactic latitude $|b|\geq25^\circ$, PSF morphology and remaining model eligibility requirements. These requirements leave \NimagingEligible systems with eligible scores for both components under at least one common model variant (either magnitude dependent or magnitude independent). At the adopted system redshift, the score $\ln\mathcal{R}$ quantifies the population-weighted photometric support for a same-redshift quasar relative to the full set of modelled explanations, including quasars at other redshifts, background sources and outliers.

We require both components to satisfy the resolved one-arcsecond search threshold of $\ln\mathcal{R}\geq$\LnRthreshold. This retains \NscoreImaging systems, comprising \NscoreImagingOne one-Gaia and \NscoreImagingTwo two-Gaia systems, shown in Figure~\ref{fig:imaging-colour-redshift}.

\begin{figure}
        \centering
        \includegraphics[width=\linewidth]{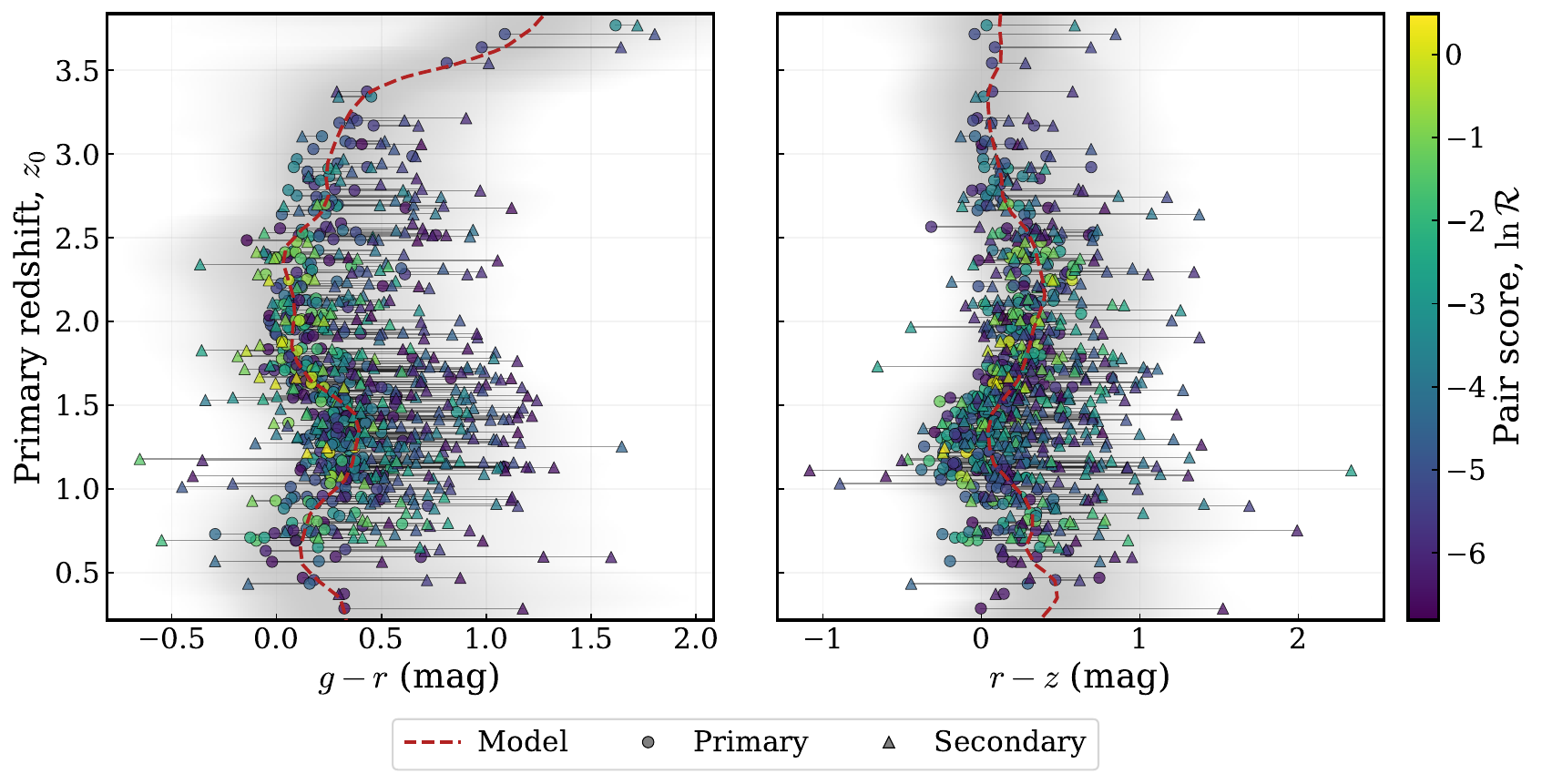}
        \caption{DECaLS $g-r$ and $r-z$ colours as a function of the adopted primary redshift for \NscoreImaging systems in the imaging pipeline satisfying $\ln\mathcal R\geq$\LnRthreshold. Circles and triangles denote primary and secondary components respectively, coloured by the system's minimum $\ln\mathcal R$ and with components belonging to the same system connected by black lines. Grey shading illustrates the quasar-model colour distribution, and the dashed red curve shows the median of the DECaLS North model.}
        \label{fig:imaging-colour-redshift}
\end{figure}

The number density distribution of the imaging candidates was analysed as a function of the component separation in Figure~\ref{fig:separation_vs_n}. As expected, we notice a significant increase in the number of candidates at higher separations, which is indicative of the selection bias towards candidates with higher separation rather than an intrinsic property of the dual AGN population itself. With future survey data releases and resolution improvements in upcoming telescopes, we might expect this distribution to extend further into the sub-kpc regime. The one-Gaia imaging route is not limited by Gaia’s ability to resolve two sources, and the final sample reaches separations of $0.177^{\prime\prime}$. The one-Gaia system J065933.20+674530.8 has a sub-kpc separation ($\Delta\theta=0.02827^{\prime\prime}$ or $r_p=0.234\,\mathrm{kpc}$ at $z=1.0322$), but its best common-model minimum component score, $\ln\mathcal R=-6.937$, fails the adopted threshold. It is therefore excluded before the pre-morphology sample. We assess the morphology of the systems that pass the score and secondary proper-motion requirements using the profile width ratio described in Section~\ref{sec:morphology}.

\begin{figure}
    \centering
    \includegraphics[width=\columnwidth]{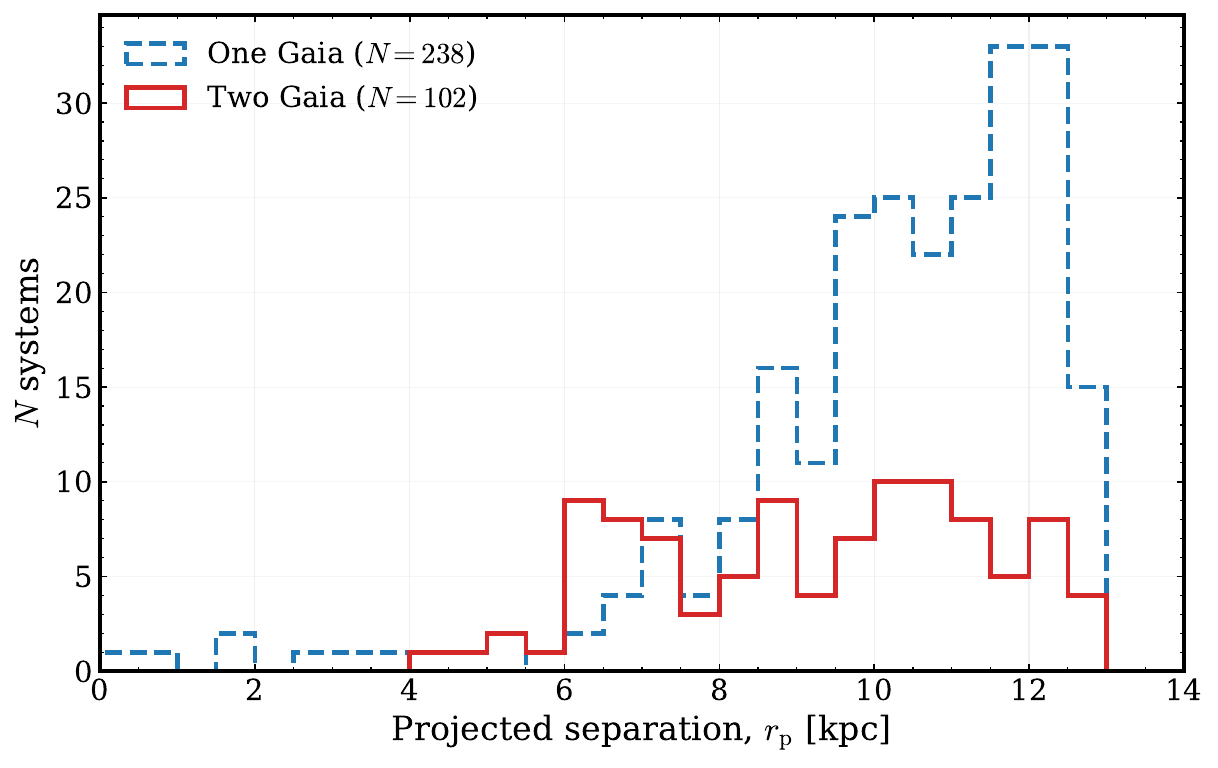}
    \caption{Distribution of projected physical separations in $0.5 \mathrm{kpc}$ bins for the \NpreMorphImaging imaging systems passing the component-score and secondary $S_{pm}$ requirements, before morphology and literature exclusions. The DCR angle cut is applied once upstream, only to one-Gaia systems. The one-Gaia and two-Gaia samples contain \NpreMorphImagingOne and \NpreMorphImagingTwo systems respectively.}
    \label{fig:separation_vs_n}
\end{figure}

Figure~\ref{fig:cutout_separation_perpendicular_along} illustrates this procedure for the two-Gaia system, J125819.31+165717.3, which has two compact quasar candidate components and a ratio of $R_\perp=1.717$. This shows how how the light distribution produces a broader profile along the separation axis for systems with multiplicity. This effect strengthens at larger separations, while light profile asymmetry along the orthogonal axes may be difficult to detect at small separations.

\begin{figure}
    \centering
    \includegraphics[width=1\linewidth]{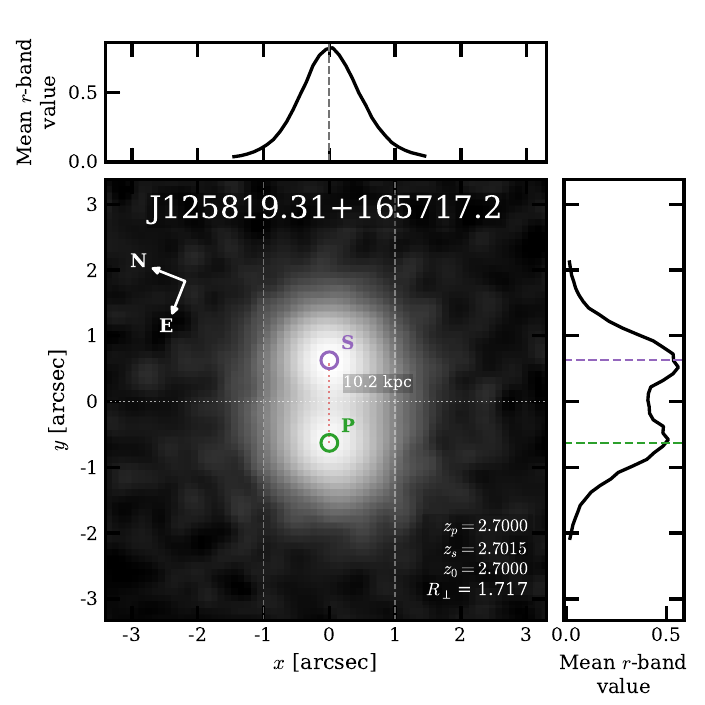}
    \caption{Example DECaLS $r$-band morphology diagnostic for the two-Gaia candidate J125819.31+165717.3. The primary and secondary components are marked by green and purple circles respectively. The dotted red line joins their positions and gives the projected physical separation of $r_p=10.2\,\mathrm{kpc}$, computed from the angular separation of $\Delta\theta=1.256^{\prime \prime}$ at the adopted system redshift $z_0=2.700$. The top and right panels show the mean $r$-band pixel value as a function of position perpendicular to and along the separation axis respectively, averaged over horizontal and vertical envelopes extending $\pm1^{\prime \prime}$ from the pair midpoint. The measured profile-width ratio, $R_{\perp}=\sigma_{\mathrm{along}}/\sigma_{\mathrm{perpendicular}}=1.717$, demonstrates that the light distribution is substantially broader along the separation axis than perpendicular to it, as expected for a resolved or semi-resolved dual source.}
    \label{fig:cutout_separation_perpendicular_along}
\end{figure}

We measure the profile-width ratio for the \NpreMorphImaging systems passing the preceding imaging selection. We apply an empirically motivated threshold of $R_\perp\geq1.05$, retaining \NpreImaging imaging candidates before the literature contaminant exclusions. The full evolution of the sample size through each stage of the imaging candidate selection pipeline is summarised in Table~\ref{tab:imaging_pcolor_pipeline_results}.
      
\begin{table}
    \centering
    \small
    \renewcommand{\arraystretch}{1.5}
    \caption{Cumulative candidate counts through the imaging pipeline
    applied to the SDSS+DESI+Quaia master catalogue.}
    \label{tab:imaging_pcolor_pipeline_results}
    \begin{tabularx}{\linewidth}{@{}Xrrr@{}}
        \toprule
        \textbf{Selection stage}
            & \textbf{One-Gaia}
            & \textbf{Two-Gaia}
            & \textbf{Total} \\
        \midrule
        Initial two-component PSF systems
            & --- & --- & 16,947 \\
        Two validated components with fixed primary assignment
            & --- & --- & 14,369 \\
        Fixed primary has Gaia
            & --- & --- & 13,803 \\
        Gaia-tier definition
            & 12,559 & 1,244 & 13,803 \\
        Primary $S_{\mathrm{PM}}<3$
            & 9,045 & 782 & 9,827 \\
        $|\Delta(g-r)_{\mathrm{s-p}}|\leq1$
            & 2,098 & 491 & 2,589 \\
        Pair--DCR angle $>20^\circ$ for One-Gaia only
            & 1,139 & 491 & 1,630 \\
        Physical-system deduplication
            & 1,139 & 461 & 1,600 \\
        Reference band $S/N\geq10$ for both components
            & 439 & 419 & 858 \\
        $|b|\geq25^\circ$ for both components
            & 390 & 351 & 741 \\
        PSF morphology for both components
            & 390 & 351 & 741 \\
        Both components eligible with finite scores in at least
        one common model variant
            & 298 & 309 & 607 \\
        $\ln\mathcal{R}\geq$\LnRthreshold for both components
            & 238 & 168 & 406 \\
        Secondary Gaia $S_{\mathrm{PM},s}<3$ for Two-Gaia;
        missing retained
            & 238 & 102 & 340 \\
        Morphology measured
            & 238 & 102 & 340 \\
        $R_\perp\geq1.05$
            & 235 & 102 & 337 \\
        After literature contaminant exclusions
            & \textbf{235} & \textbf{100} & \textbf{335} \\
        \bottomrule
    \end{tabularx}
\end{table}

The quantitative selection retains \NpreImaging imaging systems. After the literature contaminant exclusions, the final imaging sample contains \Nimaging candidates, comprising \NimagingOne one-Gaia and \NimagingTwo two-Gaia systems.

\subsection{Chance alignments}\label{sec:chance}
An unrelated star or quasar projected within $1.5^{\prime\prime}$ of a quasar can pass our selection. To measure how often this happens, we take field sources $3-10^{\prime\prime}$ from quasars in \Ncontrolfields random fields, match them in brightness, sky position and reddening to the companions of our input systems, and score each as a fake companion of a real input primary, with that primary's redshift. For blended systems we add the flux of an isolated quasar donor to the field source's flux and score the sum. The fakes are scored with the same photometric and proper-motion requirements as the real pairs. The effects of the one-Gaia DCR cut and the imaging morphology cut are included using survival factors measured from the real cutflow for each route and Galactic-latitude bin.

\begin{table}
    \centering
    \small
    \renewcommand{\arraystretch}{1.25}
    \setlength{\tabcolsep}{3pt}
    \caption{False-positive acceptance rates from unrelated controls drawn from the $3$--$10^{\prime\prime}$ annulus around quasars.
    Expected chance-alignment counts assume that every input pair is unrelated and are calculated by summing $N_{\mathrm{input},b}f_{\mathrm{pass},b}$ over Galactic-latitude bins, $b$.}
    \label{tab:chance_alignment_rates}
    \begin{tabularx}{\columnwidth}{@{}>{\raggedright\arraybackslash}Xr>{\centering\arraybackslash}p{0.24\columnwidth}>{\centering\arraybackslash}p{0.18\columnwidth}r@{}}
        \toprule
        \textbf{Route}
            & \textbf{Inputs}
            & \shortstack{\textbf{Pass rate [\%]}\\
                          \textbf{(95\% interval)}}
            & \shortstack{\textbf{Expected}\\\textbf{chance}\\
                          \textbf{alignments}}
            & \textbf{Selected} \\
        \midrule
        Gaia-neighbour: resolved
            & 309
            & \shortstack{6.89\\(5.79--8.13)}
            & 21.3 & 101 \\
        Gaia-neighbour: blended
            & 149
            & \shortstack{15.98\\(13.11--19.20)}
            & 23.6 & 40 \\
        Imaging: two-Gaia
            & 1,031
            & \shortstack{4.18\\(3.51--4.92)}
            & 43.1 & 102 \\
        Imaging: one-Gaia
            & 11,634
            & \shortstack{2.44\\(2.09--2.83)}
            & 283.4 & 235 \\
        \bottomrule
    \end{tabularx}
\end{table}

Table~\ref{tab:chance_alignment_rates} gives the pass rate for each route. If every input pair were a chance alignment, the expected number of chance alignments among the selected candidates would be the number of inputs times this rate, summed over latitude bins: \BchanceResolved, \BchanceBlended, \BchanceTwo and \BchanceOne for the resolved, blended, two-Gaia and one-Gaia routes, compared with \NpreResolved, \NpreBlended, \NpreImagingTwo and \NpreImagingOne selected. These numbers assume that every input pair is unrelated. Physically associated pairs among the inputs would reduce the true number of chance alignments, so the numbers are upper limits. Across the three routes with two Gaia detections, at most \Bmain of the \Nmain main-tier candidates are expected to be chance alignments. This sum counts the \Nshared systems selected by both searches twice, so it slightly overestimates the number. For the one-Gaia route the expected number (\BchanceOne) exceeds the number selected (\NimagingOne), so these candidates are consistent with being chance alignments, and we provide them only as targets for follow-up. We do not use the number of Gaia neighbours within $1^{\prime\prime}$ to estimate the fraction of unrelated inputs. Gaia detects fewer neighbours there than the 10--60$^{\prime\prime}$ density predicts, because it misses faint sources close to bright quasars. The fakes do not reproduce blending with a bright neighbour, so the rates are approximate.

\subsection{Properties of the final candidates}
In Table~\ref{tab:candidate_sample_summary} we present a summary of the final catalogue of dual quasar candidates from both the one arcsecond and imaging pipelines and plot their distribution of absolute galactic latitude, $|b|$, in Figure~\ref{fig:n_systems_vs_galactic_latitude_unresolved}.  We match our \NpreTotal candidates against published lensed quasars, dual AGN and known contaminants. Of these, \Nlens are confirmed lensed quasars and \NconfirmedDual are confirmed dual AGN. We also find \Nstar quasar--star alignments, \Ngal quasar--galaxy alignments and \Nproj projected quasar pairs; we remove these \Nnonlens systems. Uncertain or conflicting classifications are kept and flagged.

\begin{table*}
    \centering
    \small
    \renewcommand{\arraystretch}{1.25}
    \caption{Summary properties of the \Nfinal candidates after the adopted exclusions. The main tier contains systems with two distinct Gaia detections; the additional tier contains one-Gaia imaging systems. \Nshared systems are shared between the imaging and Gaia-neighbour searches and are counted once in the unique main tier. Literature matches refer to confirmed dual AGN using the adopted positional crossmatch.}
    \label{tab:candidate_sample_summary}
    \begin{tabularx}{0.9\linewidth}{@{}Xrrrrr@{}}
        \toprule
        \textbf{Quantity}
            & \shortstack{\textbf{Gaia neighbour}\\\textbf{(resolved)}}
            & \shortstack{\textbf{Gaia neighbour}\\\textbf{(unresolved)}}
            & \shortstack{\textbf{Imaging}\\\textbf{(two-Gaia)}}
            & \shortstack{\textbf{Main tier}\\\textbf{unique}}
            & \shortstack{\textbf{Additional tier}\\\textbf{one-Gaia}} \\
        \midrule
        Total candidates
            & \Nresolved & \Nblended & \NimagingTwo
            & \textbf{\Nmain} & \textbf{\NimagingOne} \\
        Redshift range
            & \zRangeRes & \zRangeBl & \zRangeImTwo
            & \zRangeMain & \zRangeOneGaia \\
        Median redshift
            & \zMedRes & \zMedBl & \zMedImTwo
            & \zMedMain & \zMedOneGaia \\
        Separation range [$^{\prime\prime}$]
            & \SepRangeRes & \SepRangeBl & \SepRangeImTwo
            & \SepRangeMain & \SepRangeOneGaia \\
        Projected physical separation range [kpc]
            & \kpcRangeRes & \kpcRangeBl & \kpcRangeImTwo
            & \kpcRangeMain & \kpcRangeOneGaia \\
        Median projected physical separation [kpc]
            & \kpcMedRes & \kpcMedBl & \kpcMedImTwo
            & \kpcMedMain & \kpcMedOneGaia \\
        Literature-matched candidates
            & \NlitMatchRes & \NlitMatchBl & \NlitMatchImTwo
            & \NlitMatchMain & \NlitMatchOneGaia \\
        \bottomrule
    \end{tabularx}
\end{table*}

Given that quasars are thought to be somewhat uniformly distributed across the sky, we would expect a similar result in dual quasars. When corrected for galactic latitude search area, we find that this expectation is mostly upheld in our three separate catalogues. There is, however, an apparent underdensity of systems from the imaging pipeline below $|b|\lesssim40^\circ$, which may result from overly conservative quality thresholds removing candidates at lower galactic latitudes. Additionally, there appears to be a slight increase in the number of systems from the unresolved one-arcsecond search toward lower galactic latitude. While this may arise from a small sample, survey coverage or the distribution of parent quasars, it is also likely that the difficulty in both detecting and removing systems contaminated with unresolved, projected stellar companions at lower galactic latitudes impacts the counts in this pipeline. As such, high resolution follow-up imaging or spectroscopy of these candidates would be required to determine the nature of the systems.

\begin{figure}
    \centering
    \includegraphics[width=1\linewidth]{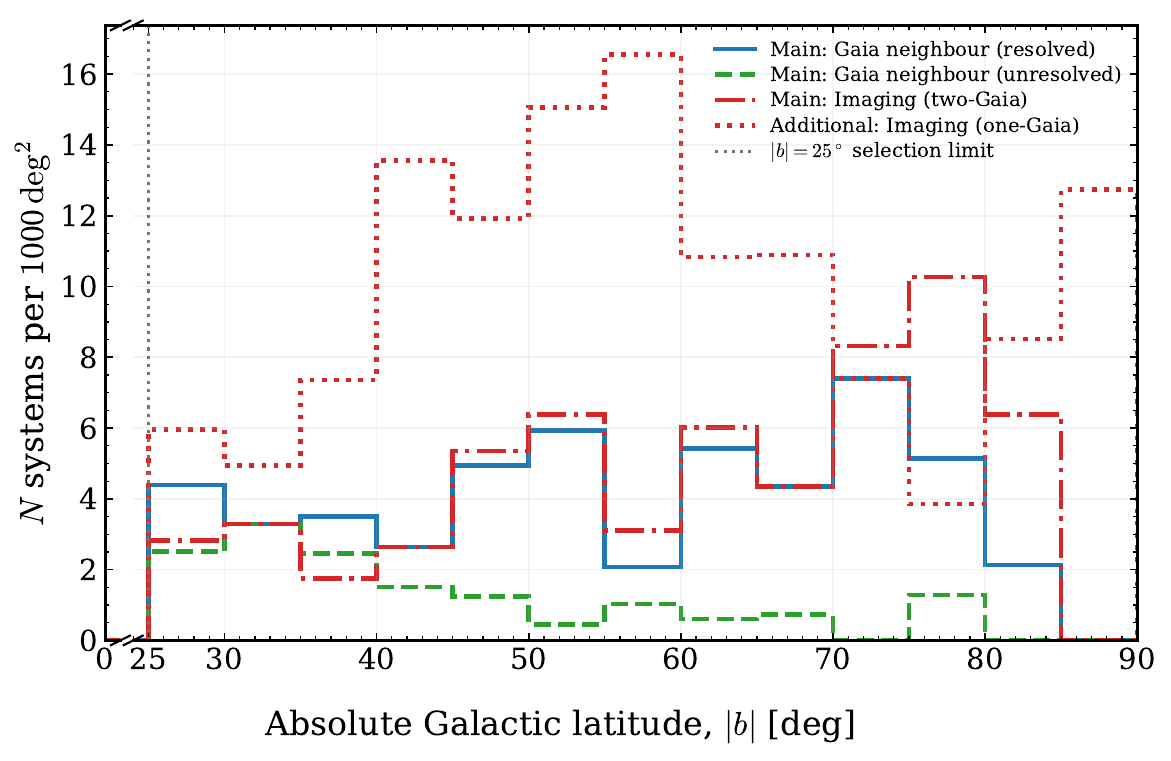}
    \caption{Distribution of candidate systems in $5^\circ$ bins of absolute Galactic latitude, $|b|$, normalised by each bin’s geometric sky area. The blue and green lines show the resolved and unresolved candidates from the one arcsecond search respectively while the red line shows the candidates from the imaging search. The vertical line marks the $|b|=25^\circ$ selection limit.}
    \label{fig:n_systems_vs_galactic_latitude_unresolved}
\end{figure}

 We match the \NpreOneArcsec one arcsecond candidates (\NpreResolved resolved, \NpreBlended blended) and the \NpreImaging imaging candidates within $0.1^{\prime\prime}$, allowing the primary and secondary labels to swap. The two searches share \NpreShared systems, giving \NpreTotal unique candidates before the literature step.

We crossmatch our \NpreTotal candidates against \NlitDuals confirmed dual AGN from the literature with $0 \leq z \leq 4$ and angular separations $\leq 1.5^{\prime\prime}$. We search for a published position within $1^{\prime\prime}$ of either candidate component, with \NconfirmedDual of our candidates matching previously confirmed dual AGN. We verify \NlitPairVerified matches by requiring both components to agree with their published positions to within $0.1^{\prime\prime}$, and confirm \NlitGeometryVerified further matches by reviewing the published evidence for the identity of each system. Of the \NlitUnmatched confirmed literature dual AGN without a match in our candidates, \NlitNotInMaster are absent from the master quasar catalogue, with some present in SDSS or Gaia but absent from SDSS DR16Q or Quaia respectively. A further \NlitNoSecondary have no secondary object to score within $1.5^{\prime\prime}$ in DECaLS or $1^{\prime\prime}$ in Gaia, while \NlitFailPSF fail the PSF requirements (e.g. J2332+2243 is in the master catalogue and has a Tractor REX + PSF pair so fails on the imaging pipeline's PSF requirements). The remaining \NlitFailCuts systems reached pair validation or scoring but failed specific selection cuts in DCR, proper motion significance or colour scoring. Of the \NlitUnmatched unmatched systems, \NlitIndependent were originally identified through routes independent of SDSS, DESI or Gaia, and \NlitIndependentImaged of these had their two nuclei revealed in targeted high-resolution optical or near infrared images.

\begin{figure*}
    \centering
    \includegraphics[width=1\linewidth]{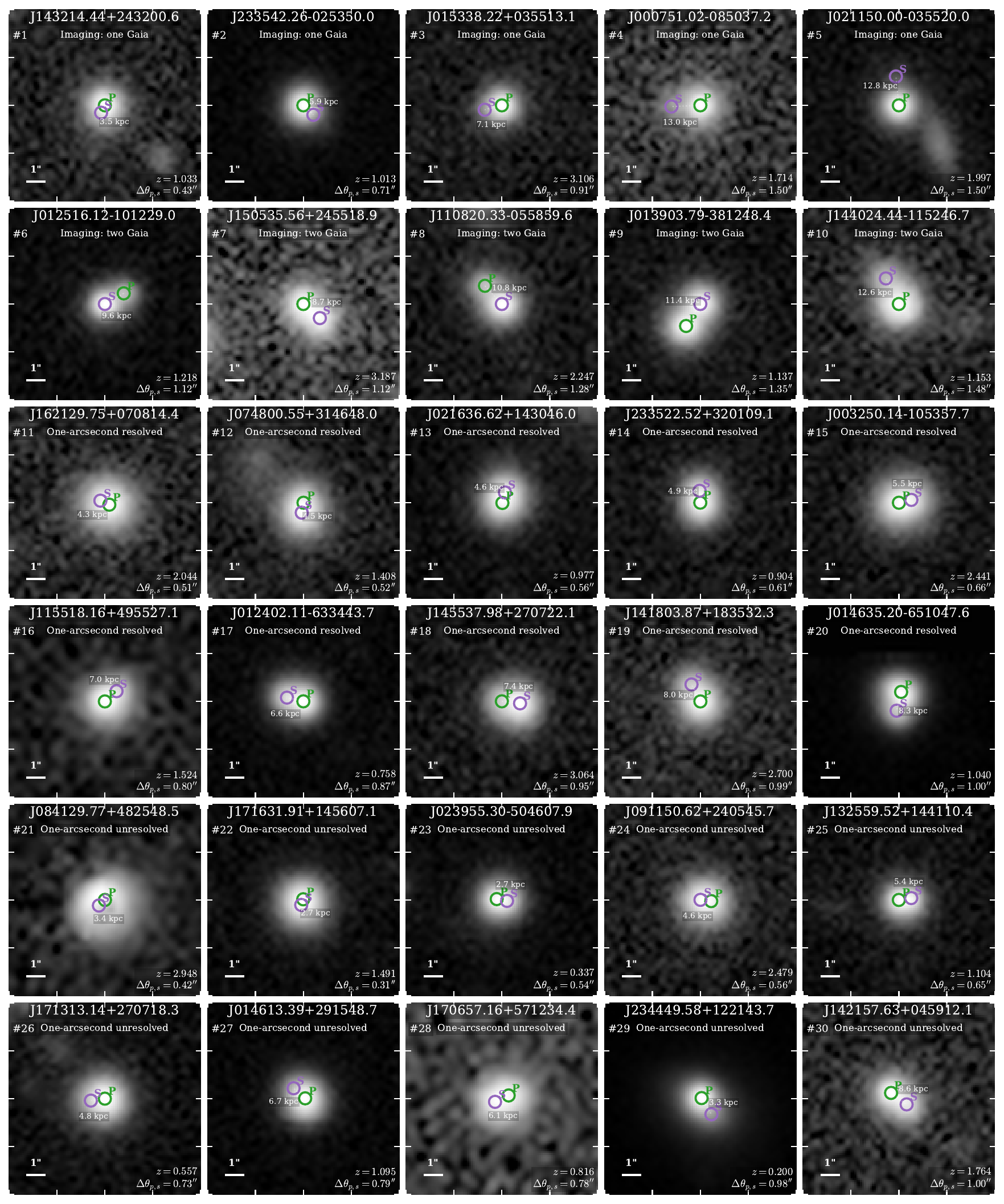}
    \caption{Representative examples of 30 candidate dual quasar systems drawn from the final \Nfinal candidates from both the imaging and one arcsecond pipelines. The first two rows show 10 imaging candidates, the next two rows 10 resolved one arcsecond candidates, and the final two rows 10 unresolved one arcsecond candidates. Primary and secondary components are marked in green and purple respectively and labelled with the systems' projected physical separation. Each panel also gives the angular separation and either the adopted system redshift (defined according to the redshift hierarchy of DESI>SDSS>Quaia) or, where available, the individual component redshifts. The first row shows five one-Gaia systems, and the remaining five rows show 25 two-Gaia systems. Images are DECaLS $r-\mathrm{band}$ cutouts.}
    \label{fig:final_candidate_cutouts}
\end{figure*}

In Figure~\ref{fig:final_candidate_cutouts}, we present a representative sample of the \Nfinal dual and lensed quasar candidates from both the imaging and one arcsecond pipelines in DECaLS  $r-\mathrm{band}$ and the distribution of projected separation as a function of redshift for the candidate systems identified in this work in Figure~\ref{fig:literature_vs_projected_separation_redshift}. We compute the projected separations as $r_{p}=\Delta\theta D_A(z_0)$, where $\Delta\theta$ is the angular separation between the two components, $z_0$ is the adopted redshift of the system and $D_A$ is the angular diameter distance. These separations assume that both components lie at $z_0$. The dotted grey lines mark fixed angular separations of $1.5^{\prime \prime}$, $1^{\prime \prime}$, $0.5^{\prime \prime}$ and $0.3^{\prime \prime}$.

To put this work into the wider context, we compare our candidates with known sub 1.5 arcsecond dual quasars/AGN from the literature, shown as grey markers. The literature sample is primarily drawn from the VODKA collaboration \citep{shenVarstrometryOffnucleusDual2019} and the BigMAC \citep{pfeifleSupersizeMeBig2025}, and GMP searches by \citet{mannucciGMPselectedDualLensed2023} as well as sources from radio and high resolution imaging observations. Of the \NpreTotal unique candidate systems from both pipelines, \NconfirmedDual have previously been confirmed as dual AGN and \Nlens are known lensed quasars.

\begin{figure*}
    \centering
    \includegraphics[width=\linewidth]{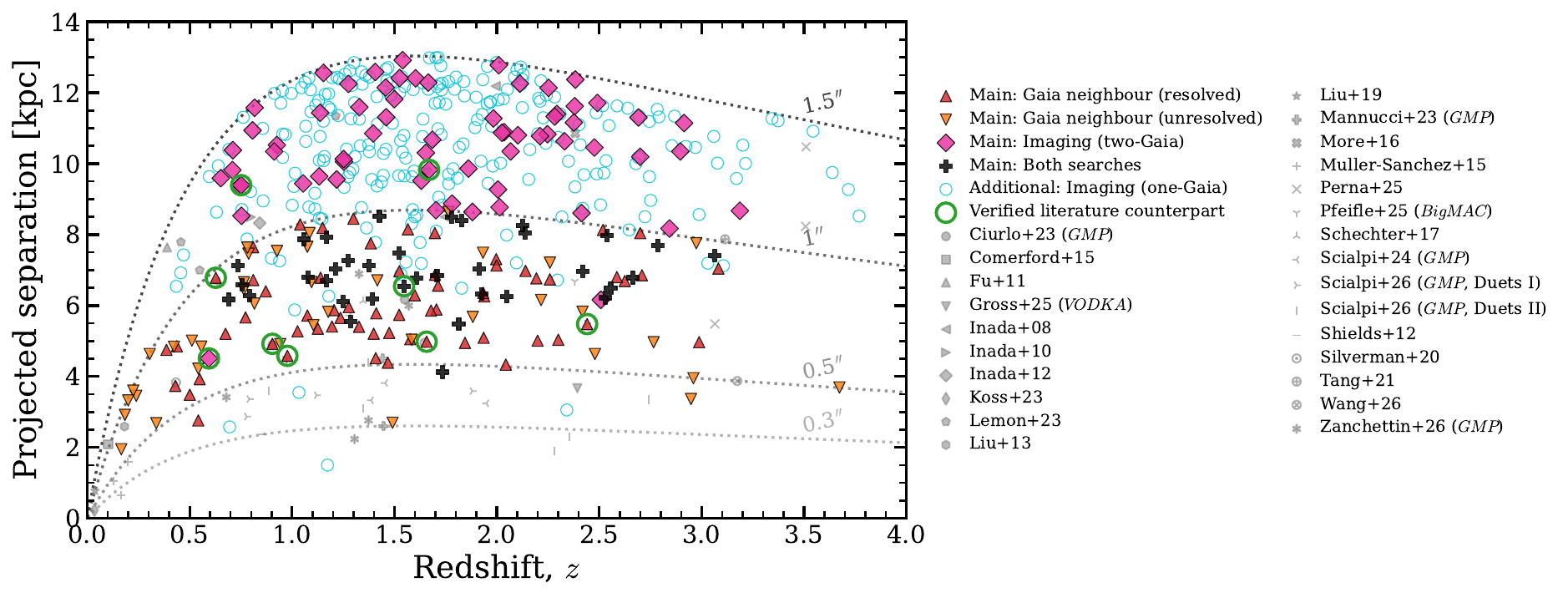}
    \caption{Projected separation versus redshift of the \Nfinal unique candidates from the Gaia one arcsecond and imaging pipelines, compared with \NlitDuals confirmed literature dual AGN at angular separations $\leq1.5^{\prime\prime}$. Coloured symbols distinguish the candidate selection routes, while grey symbols show literature systems with a distinct marker for each source. Green rings identify the \NconfirmedDual candidates matched to confirmed literature duals. Each literature system is plotted once. Dotted curves indicate angular separations of $0.3^{\prime\prime}$, $0.5^{\prime\prime}$, $1.0^{\prime\prime}$ and $1.5^{\prime\prime}$. Literature data are compiled from \citet{inadaSloanDigitalSky2008}, \citet{inadaSloanDigitalSky2010}, \citet{fuKiloparsecScaleBinaryActive2011}, \citet{inadaSloanDigitalSky2012}, \citet{shieldsLBQS01032753Binary2011}, \citet{liuChandraXrayHubble2012}, \citet{comerfordMergerdrivenFuelingActive2015}, \citet{muller-sanchezORIGINDOUBLEPEAKEDNARROW2015}, \citet{moreSDSSIIIBOSSQuasar2016}, \citet{schechterFirstLensedQuasar2017}, \citet{liuEllipticalGalaxyMaking2019}, \citet{silvermanDualSupermassiveBlack2020}, \citet{tangOpticalSpectroscopyDual2021}, \citet{ciurloNewMultipleAGN2023}, \citet{kossUGC4211Confirmed2023}, \citet{lemonGravitationallyLensedQuasars2023}, \citet{mannucciGMPselectedDualLensed2023}, \citet{scialpiMUSEAdaptiveopticsSpectroscopy2024}, \citet{grossVarstrometryOffnucleusDual2024}, \citet{pernaGANIFSHighNumber2025}, \citet{pfeifleSupersizeMeBig2025}, \citet{scialpiCosmicDuetsHighspatial2026}, \citet{scialpiCosmicDuetsII2026}, \citet{wangHSTConfirmsSub52026}, and \citet{zanchettinDualLensedAGN2026}.
    }
    \label{fig:literature_vs_projected_separation_redshift}

\end{figure*}

The systems we identify are candidates for high-resolution follow-up observations, and in most cases we cannot classify both the multiplicity and object type with absolute certainty in its current implementation. However, the recent release of Euclid DR1 \citep{collaborationEuclidFirstData2026} has provided high-resolution observation of a subset of our candidates. We therefore validate the efficacy of our pipelines in identifying multiplicity in quasar systems using this data. Among the \Nfinal candidates, \NeuclidCovered are covered by Euclid DR1, \NeuclidVIS of which have usable data from the visible instrument (VIS). Inspecting images in bands from VIS and the Near-Infrared Spectrometer and Photometer (NISP), we find that \NeuclidTwo are clear dual objects, with a further \NeuclidBlend identified as a potential blended system. Euclid confirms that these systems contain two sources, and so rules out deblending and DCR artefacts for them. It does not show whether the second source is a quasar. Of the \NeuclidVIS, \NeuclidResolved are resolved one arcsecond, \NeuclidBlended blended and \NeuclidImaging imaging candidates. These route counts include \NeuclidShared systems shared by the one arcsecond and imaging searches. Of the \NeuclidVIS systems with usable VIS data, \NeuclidMain are in the main tier and \NeuclidAdditional are in the additional tier. The system that Euclid does not clearly resolve is J094112.53+621719.4. In Figure~\ref{fig:euclid_vs_decals}, we compare DECaLS DR9, Euclid DR1 VIS and Euclid DR1 NISP cutouts for a one-Gaia imaging candidate and a candidate selected by the resolved Gaia-neighbour pipeline.

\begin{figure}
    \centering
    \includegraphics[width=1\linewidth]{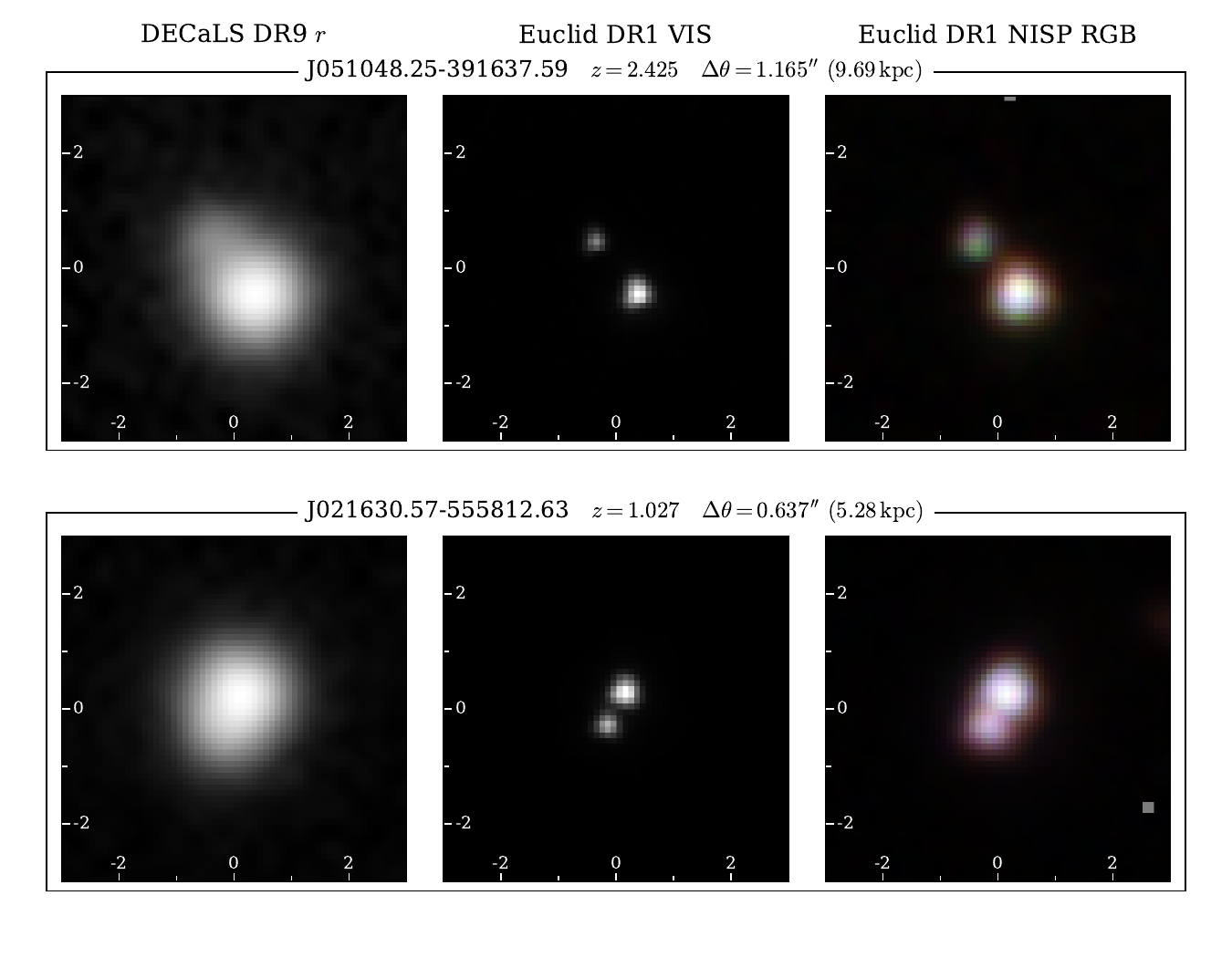}
    \caption{DECaLS DR9 $r$-band (\textbf{left}), Euclid DR1 VIS $I_{\mathrm E}$-band (\textbf{centre}), and Euclid DR1 NISP colour (\textbf{right}) cutouts. The upper row shows the one-Gaia imaging candidate J051048.25$-$391637.6 ($z=2.425$, $\Delta\theta=1.165^{\prime\prime}$). The lower row shows J021630.57$-$555812.6 ($z=1.027$, $\Delta\theta=0.637^{\prime\prime}$), selected by the resolved Gaia-neighbour pipeline. Euclid VIS separates two components in each system. NISP colours map $H_{\mathrm E}$, $J_{\mathrm E}$ and $Y_{\mathrm E}$ to red, green and blue. Each panel spans $6^{\prime\prime}\times6^{\prime\prime}$, with north up, east left and white ticks every $1^{\prime\prime}$.}
    \label{fig:euclid_vs_decals}
\end{figure}

The improvement in resolution from Euclid DR1 clearly distinguishes two distinct sources, validating our pipelines' ability to detect multiplicity near known quasars. However, in a small number of candidates, the components appear to exhibit different colours in NISP cutouts. This is unexpected of two resolved quasars at the same redshift and further photometric analysis is required to classify the nature of the secondary component. Such analysis represents the next steps in a true wide field dual AGN detection pipeline but is beyond the scope of this paper.

\section{Conclusions}
We have presented two complementary searches for close-separation dual AGN candidates using Gaia DR3 source multiplicity and ground based DECaLS imaging. Both searches use probabilistic photometric models to assess colour consistency with known quasar populations across redshifts. We evaluate independent components at the adopted primary redshift, while unresolved systems lacking component photometry are assessed with hypotheses of either two quasars or a quasar and a background companion.

The imaging search selects \Nimaging candidates, \NimagingOne of them with one Gaia detection. The one arcsecond search selects \Nresolved resolved and \Nblended blended candidates. The two searches share \Nshared systems, giving \Nfinal unique candidates after removing \Nnonlens known non-lens contaminants and \Nartefact corrupted-data artefact identified during visual inspection. Of these, \NconfirmedDual are confirmed dual AGN and \Nlens are known lensed quasars. \Nmain candidates form the main tier, with two distinct Gaia detections, and \NimagingOne form the additional tier, with one Gaia detection. All confirmed dual AGN and known lenses are in the main tier. From random-field controls we expect at most \Bmain chance alignments in the main tier, while the additional tier is consistent with chance alignments and is provided for follow-up. The candidates have angular separations spanning \SepRange and adopted redshifts spanning \zRange. Assuming that both components lie at the adopted system redshift, these correspond to projected separations of \kpcRange.

The complementary selections allow us to investigate systems with different available component data, including candidates absent from a search requiring two Gaia detections and those without sufficient resolved component photometry. However, consistency between our candidates and quasar photometry does not necessarily establish physical pairs and gravitationally lensed quasars, projected stellar companions and source modelling artefacts remain possible explanations for observed candidates. The sample should therefore be interpreted as a catalogue for further follow-up observations rather than a confirmed population of dual AGN.

High-resolution imaging and spatially resolved spectroscopy are required to determine the nature of these systems and measure the fraction containing two physically associated active nuclei. Such observations would allow us to qualify both methodologies and their efficacy in identifying true dual AGN as well as enable further studies of SMBH pairing during galaxy mergers and help connect the observable sub-arcsecond population to the formation of compact SMBH binaries.

\section*{Acknowledgements}

LC acknowledges support from the Science and Technology Facilities Council (STFC). This work uses data from ESA's Gaia mission, alongside products from the DESI Legacy Imaging Surveys, SDSS-IV, DESI DR1, Quaia and Euclid DR1. We acknowledge the survey teams, ESA, the Euclid Consortium, Gaia Data Processing and Analysis Consortium and the participating institutions and funding agencies that supported these missions.

\section*{Data Availability}
The final catalogue of \Nfinal candidates is provided as two machine-readable tables, one containing \Nmain main-tier systems and the other containing \NimagingOne additional-tier systems. The complete tables, their combined catalogue and column definitions are available in the \texttt{catalogue/} directory at \url{https://github.com/lewis-cooper/hunting-dual-agn} and accompany the paper as ancillary material. Example entries and column descriptions are given in Appendix~\ref{app:catalogues}. These publication tables exclude the corrupted-data artefact identified during visual inspection; the automatic pipeline output is retained separately. The survey data are available from the respective survey archives.

Open AI's ChatGPT and Codex were used to assist in the development and optimisation of code, plot formatting, manuscript editing and grammar checking. The authors retain full responsibility for the content of the paper.

\bibliographystyle{mnras}
\bibliography{references}

\appendix

\section{Candidate catalogues}
\label{app:catalogues}

The complete electronic catalogue is supplied in two parts: the main tier
contains \Nmain systems with two distinct Gaia detections, and the additional tier contains \NimagingOne one-Gaia imaging candidates. Tables~\ref{tab:catalogue_main} and~\ref{tab:catalogue_additional} show the first ten entries of each part, sorted by primary right ascension. Table~\ref{tab:catalogue_columns} describes all columns in the machine-readable files. Every system's J-coordinate name is based on its adopted primary position. Numeric coordinates preserve the precision of the analysis catalogue. A shared redshift is assumed when calculating projected separations and does not establish a physical association.

\input{tables/catalogue_main_preview.tex}
\input{tables/catalogue_additional_preview.tex}
\input{tables/catalogue_columns.tex}

\bsp	%
\label{lastpage}
\end{document}

%% file: tables/catalogue_main_preview.tex
\begin{table*}
\centering
\footnotesize
\setlength{\tabcolsep}{4pt}
\caption{Example entries from the main tier. The complete machine-readable table contains \Nmain systems. Column definitions are given in Table~\ref{tab:catalogue_columns}. Route codes are R (resolved Gaia neighbour), U (unresolved Gaia neighbour), I1 (one-Gaia imaging) and I2 (two-Gaia imaging); a plus sign marks selection by both searches.}
\label{tab:catalogue_main}
\begin{tabular}{@{}lclrrrrrl@{}}
\toprule
Name & Tier & Route & RA & Dec & $\Delta\theta$ & $z$ & $r_p$ & Literature class \\
 & & & [deg] & [deg] & [$^{\prime\prime}$] & & [kpc] & \\
\midrule
J000251.21+130915.8 & main & R & 0.713355 & +13.154378 & 0.727 & 1.939 & 6.26 & Unclassified \\
J002348.32+230935.5 & main & R+I2 & 5.951348 & +23.159863 & 0.956 & 2.785 & 7.70 & Unclassified \\
J002942.86-694729.7 & main & U & 7.428578 & -69.791594 & 0.971 & 0.788 & 7.46 & Unclassified \\
J003250.14-105357.7 & main & R & 8.208900 & -10.899358 & 0.659 & 2.441 & 5.48 & Dual AGN \\
J003900.90+164345.6 & main & R & 9.753749 & +16.729341 & 0.897 & 1.384 & 7.75 & Unclassified \\
J004127.46-020332.4 & main & R & 10.364402 & -2.059001 & 0.573 & 1.844 & 4.95 & Unclassified \\
J004527.94-393734.0 & main & I2 & 11.366402 & -39.626124 & 1.142 & 1.861 & 9.86 & Lens \\
J004929.80-725550.4 & main & U & 12.374183 & -72.930653 & 0.997 & 0.305 & 4.63 & Unclassified \\
J005027.84-174010.6 & main & I2 & 12.615999 & -17.669605 & 1.428 & 1.603 & 12.41 & Lens \\
J005717.08-155154.6 & main & R & 14.321149 & -15.865165 & 0.819 & 2.625 & 6.69 & Unclassified \\
\bottomrule
\end{tabular}
\end{table*}

%% file: tables/catalogue_additional_preview.tex
\begin{table*}
\centering
\footnotesize
\setlength{\tabcolsep}{4pt}
\caption{Example entries from the additional tier. The complete machine-readable table contains \NimagingOne systems. Column definitions are given in Table~\ref{tab:catalogue_columns}. Route codes are R (resolved Gaia neighbour), U (unresolved Gaia neighbour), I1 (one-Gaia imaging) and I2 (two-Gaia imaging); a plus sign marks selection by both searches.}
\label{tab:catalogue_additional}
\begin{tabular}{@{}lclrrrrrl@{}}
\toprule
Name & Tier & Route & RA & Dec & $\Delta\theta$ & $z$ & $r_p$ & Literature class \\
 & & & [deg] & [deg] & [$^{\prime\prime}$] & & [kpc] & \\
\midrule
J000751.02-085037.2 & additional & I1 & 1.962578 & -8.843654 & 1.497 & 1.714 & 12.99 & Unclassified \\
J001846.89-514420.0 & additional & I1 & 4.695379 & -51.738891 & 1.196 & 1.377 & 10.34 & Unclassified \\
J002215.36+000121.6 & additional & I1 & 5.563981 & +0.022673 & 0.983 & 1.608 & 8.54 & Unclassified \\
J002304.44+000515.0 & additional & I1 & 5.768481 & +0.087506 & 1.108 & 1.807 & 9.59 & Unclassified \\
J002553.80+183255.1 & additional & I1 & 6.474175 & +18.548640 & 1.124 & 0.435 & 6.54 & Unclassified \\
J002615.77-410824.1 & additional & I1 & 6.565729 & -41.140028 & 1.284 & 1.417 & 11.12 & Unclassified \\
J002926.22+112249.5 & additional & I1 & 7.359253 & +11.380427 & 1.230 & 2.679 & 10.00 & Unclassified \\
J003156.72+180043.0 & additional & I1 & 7.986353 & +18.011953 & 1.392 & 1.730 & 12.08 & Unclassified \\
J003210.77+224303.1 & additional & I1 & 8.044878 & +22.717533 & 1.432 & 2.783 & 11.53 & Unclassified \\
J003530.92-650600.9 & additional & I1 & 8.878829 & -65.100250 & 1.402 & 1.708 & 12.17 & Unclassified \\
\bottomrule
\end{tabular}
\end{table*}

%% file: tables/catalogue_columns.tex
\begin{table*}
\centering
\footnotesize
\caption{Columns in the machine-readable candidate catalogues. Blank fields indicate unavailable measurements or inapplicable routes. Names use the primary position; the numeric coordinates retain their original precision. Gaia identifiers must be read as text.}
\label{tab:catalogue_columns}
\begin{tabularx}{\linewidth}{@{}llX@{}}
\toprule
Column & Unit & Description \\
\midrule
candidate\_id & --- & Stable identifier linking the system to the automatic score audits. \\
system\_name & --- & J-coordinate name from the adopted primary ICRS position; RA seconds rounded to two decimals and Dec seconds to one. \\
original\_system\_name & --- & Unmodified name or native catalogue identifier in the automatic catalogue. \\
tier & --- & main: two distinct Gaia detections; additional: one-Gaia imaging selection. \\
selection\_route & --- & All selecting routes, separated by semicolons: gaia\_resolved, gaia\_unresolved, imaging\_one\_gaia or imaging\_two\_gaia. \\
primary\_ra\_deg & deg & Adopted primary right ascension, ICRS. \\
primary\_dec\_deg & deg & Adopted primary declination, ICRS. \\
secondary\_ra\_deg & deg & Adopted secondary right ascension, ICRS. \\
secondary\_dec\_deg & deg & Adopted secondary declination, ICRS. \\
separation\_arcsec & arcsec & Angular component separation from the automatic catalogue. \\
redshift & --- & Adopted system redshift; a shared component redshift is an assumption, not confirmation. \\
redshift\_catalogue & --- & Catalogue supplying the adopted redshift. \\
adopted\_redshift\_error\_used & --- & Additional redshift uncertainty used in scoring; zero for spectroscopic redshifts. \\
redshift\_treatment & --- & Fixed spectroscopic redshift or marginalisation over a photometric redshift uncertainty. \\
projected\_separation\_kpc & kpc & Projected separation assuming both components share the adopted redshift; Planck18 cosmology. \\
primary\_gaia\_source\_id & --- & Exact Gaia DR3 identifier of the primary, stored as text. \\
secondary\_gaia\_source\_id & --- & Exact Gaia DR3 identifier of the secondary; blank when absent. \\
selected\_by\_resolved & --- & True when selected by the resolved Gaia-neighbour route. \\
selected\_by\_unresolved & --- & True when selected by the unresolved Gaia-neighbour route. \\
selected\_by\_imaging & --- & True when selected by the imaging search. \\
selected\_by\_both\_searches & --- & True when selected by both Gaia-neighbour and imaging searches. \\
imaging\_selection\_route & --- & one\_gaia or two\_gaia for imaging selections; blank otherwise. \\
secondary\_pm\_significance & --- & Secondary Gaia proper-motion significance, when measured. \\
secondary\_pm\_missing & --- & True when a two-Gaia system has no usable secondary proper-motion significance. \\
n\_psf\_components\_within\_aperture & --- & Number of Tractor PSF sources inside the imaging search aperture, when available. \\
literature\_class & --- & dual\_agn, lens, uncertain or unclassified; literature identifications are not new confirmations. \\
literature\_reference\_ids & --- & Identifiers of the matching records in the literature comparison catalogues. \\
literature\_author\_year & --- & Author/year labels for literature classifications. \\
literature\_publication\_urls & --- & Source-paper links for literature classifications. \\
\bottomrule
\end{tabularx}
\end{table*}